\documentclass[sigconf]{acmart}
\AtBeginDocument{%
  }

\copyrightyear{2026}
\acmYear{2026}
\setcopyright{cc}
\setcctype{by-nc-nd}
\acmConference[CHI '26]{Proceedings of the 2026 CHI Conference on Human Factors in Computing Systems}{April 13--17, 2026}{Barcelona, Spain}
\acmBooktitle{Proceedings of the 2026 CHI Conference on Human Factors in Computing Systems (CHI '26), April 13--17, 2026, Barcelona, Spain}
\acmPrice{}
\acmDOI{10.1145/3772318.3790396}
\acmISBN{979-8-4007-2278-3/2026/04}

\usepackage{multirow}
\usepackage{color}
\usepackage{graphicx}
\usepackage{float}
\usepackage{subfigure}
\usepackage{verbatim}  
\usepackage{enumitem}
\usepackage{makecell}
\begin{document}


\title{DroidRetriever: A Transparent and Steerable Automation System for Collaborative Mobile Information Seeking}


\author{Yiheng Bian}
\affiliation{
  \department{MOE KLINNS Lab}
  \institution{Xi'an Jiaotong University}
  \city{Xi'an}
  \country{China}
}
\email{yhbian@stu.xjtu.edu.cn}

\author{Yunpeng Song}
\affiliation{
  \department{MOE KLINNS Lab}
  \institution{Xi'an Jiaotong University}
  \city{Xi'an}
  \country{China}
}
\email{yunpengs@xjtu.edu.cn}
\authornote{Corresponding author.}

\author{Guiyu Ma}
\affiliation{
  \department{MOE KLINNS Lab}
  \institution{Xi'an Jiaotong University}
  \city{Xi'an}
  \country{China}
}
\email{guiyu.ma@stu.xjtu.edu.cn}

\author{Rongrong Zhu}
\affiliation{
  \department{MOE KLINNS Lab}
  \institution{Xi'an Jiaotong University}
  \city{Xi'an}
  \country{China}
}
\email{zhurongorng@stu.xjtu.edu.cn}

\author{Zhongmin Cai}
\affiliation{
  \department{MOE KLINNS Lab}
  \institution{Xi'an Jiaotong University}
  \city{Xi'an}
  \country{China}
}
\email{zmcai@sei.xjtu.edu.cn}
\authornotemark[1]

\renewcommand{\shortauthors}{Bian et al.}

\begin{abstract}

Information seeking on mobile devices is often fragmented, trapping users in repetitive cycles of context switching and data re-entry, which increases cognitive load and disrupts workflow. Existing mobile agents provide limited cross-source integration and are largely opaque, presenting progress as a linear feed with few opportunities to intervene, steer, or take control. We present DroidRetriever, a transparent, steerable system for cross-source mobile information seeking. It accepts voice or typed input and the multi-LLM system decomposes the task, navigates to target pages, takes screenshots, and synthesizes a concise report with citation-linked screenshots. We make the process transparent through a progress dashboard combining sub-task progress and real-time exploration maps for seamless takeover. DroidRetriever also pauses on detected privacy or high-risk screens and prompts intervention. Across 35 tasks over 24 apps, experiments and user studies demonstrate improvements in coverage, transparency, and reduced workload. \textcolor{black}{We release our code at https://github.com/AkimotoAyako/DroidRetriever.}

\end{abstract}




\maketitle

\section{Introduction}

Mobile ecosystems provide a vast array of domain-specific services, allowing users to access vertically integrated solutions for needs ranging from travel planning to price comparison. While these services are highly optimized for their specific verticals, real-world user goals often transcend the boundaries of any single provider. Users frequently need to orchestrate workflows across multiple services, gathering and evaluating information in real time. However, current mobile interaction models are constrained by rigid service encapsulations, where data and interactions are confined within isolated environments. This structural fragmentation forces users to bridge the gap manually: instead of focusing on their primary goals, they are compelled to engage in repetitive cycles of context switching and information foraging across disjointed interfaces~\cite{Pirolli1995, Pirolli2005}.

\begin{figure*}[]
    \captionsetup{justification=justified, singlelinecheck=false}
    \Description[Comparison between DroidRetriever (left) and the general-purpose LLM-driven agent (right).]{Figure 1: illustrates Comparison between DroidRetriever (left) and the general-purpose LLM-driven agent (right). DroidRetriever consists of three phases: Task Decomposition, UI Copilot (and Intervention), and Report Synthesis. Users interact with the system by describing tasks in natural language (text or speech), checking task progress through the dashboard, pausing to take over at any time or when prompted by phone vibration, and reviewing the final results at the bottom of the dashboard.}
    \centering
    \includegraphics[width=1\linewidth]{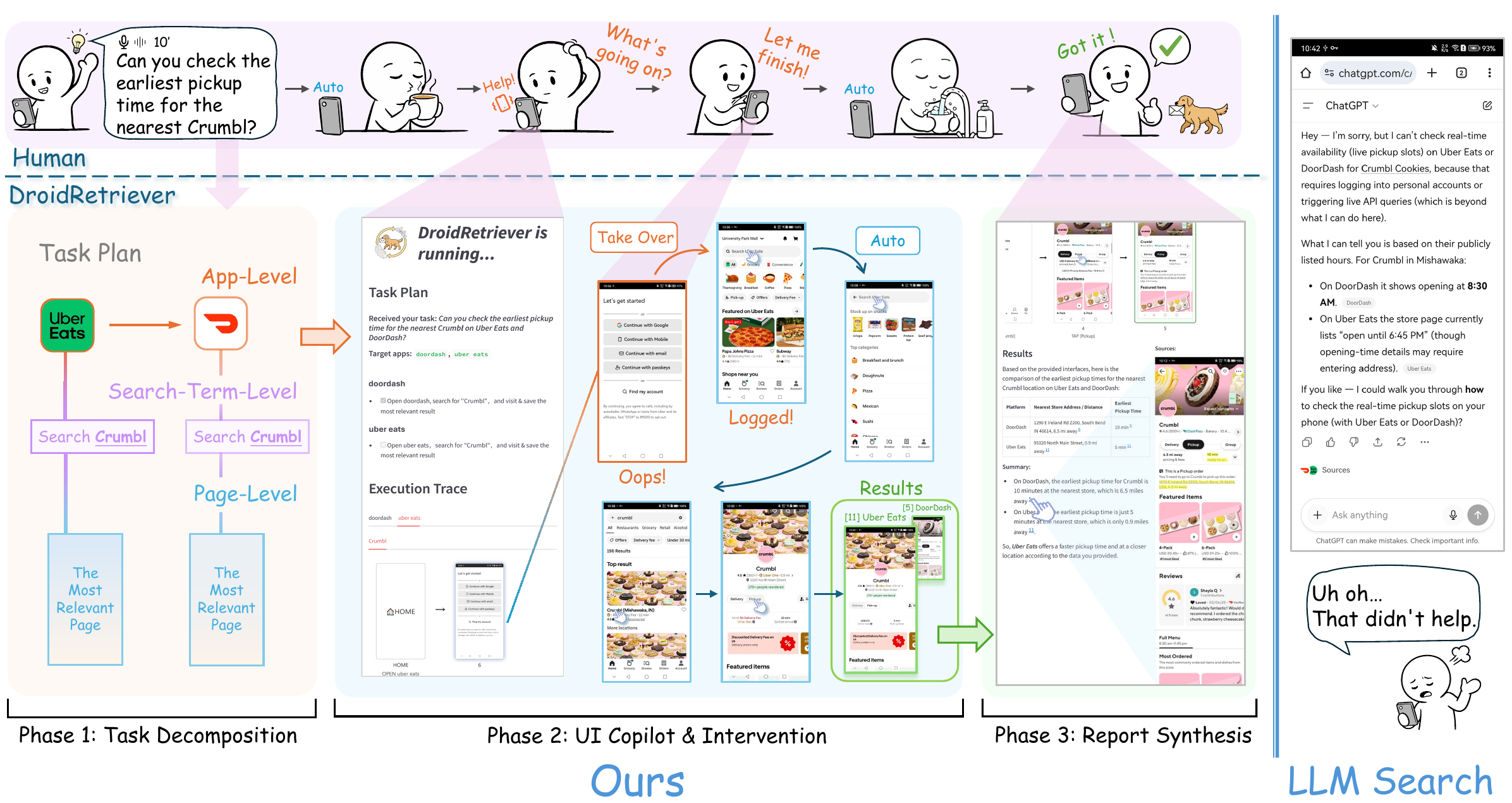}
    \caption{Comparison between DroidRetriever (left) and the general-purpose LLM-driven agent (right). DroidRetriever consists of three phases: Task Decomposition, UI Copilot (and Intervention), and Report Synthesis. Users interact with the system by describing tasks in natural language (text or speech), checking task progress through the dashboard, pausing to take over at any time or when prompted by phone vibration, and reviewing the final results at the bottom of the dashboard.}
    \label{fig:all-frame}
\end{figure*}


This fragmentation imposes a significant interaction overhead, characterized by both temporal and spatial dimensions. Temporally, the workflow is disrupted by the high cost of navigation. Users must repeatedly traverse app launchers, perform multi-step searches, and deep navigation stacks to locate specific items. These sequential operations break the continuity of the task, making it difficult to maintain focus, especially in distracted mobile scenarios~\cite{chang2016}. Spatially, the lack of a unified information workspace creates barriers to information synthesis. Unlike desktop environments that support parallel viewing, mobile users must toggle between separate screens to access distributed content, such as product specifications or policy details~\cite{HOLLENDER20101278}. This separation forces users to rely heavily on working memory to connect fragmented data, a process further hindered by UI limitations like the inability to select or copy text from images easily~\cite{Gupta16, Bento2018}.


Various tools have been developed to enhance the information-seeking process on mobile devices, which can be broadly classified into two categories. The first category accesses app data by direct API calls, much like Siri retrieves calendar information~\cite{GPT-Retrieval-Plugin, Perplexica, Morphic}. However, this method's dependence on a fixed set of predefined APIs makes it inflexible for tasks involving unsupported sources.The second category consists of general-purpose LLM-driven search agents, such as ChatGPT and Perplexity AI~\cite{SearchGPT, Lepton}. These agents typically translate the user's task into search queries, retrieve relevant information using a search engine, and employ the LLM to synthesize the findings into a coherent summary. However, this approach is constrained in both login-gated and open-web scenarios. On login-gated platforms, the requirement for user authentication makes accessing dynamic in-app content a primary obstacle for autonomous agents. Even on platforms that are seemingly part of the open web, these agents face significant hurdles. For instance, lacking user context, such as login credentials, they often retrieve generic data, rendering context-sensitive information like pricing inaccurate. Furthermore, to compensate for retrieval failures from specified sources, they may silently fallback to alternative platforms, generating plausible yet erroneous answers that require manual verification. As shown on the right of Figure~\ref{fig:all-frame}, the LLM fails to retrieve a specific pickup schedule, offering only statewide business hours due to its inability to interact with the currently available web elements.

In this paper, we propose DroidRetriever, a transparent, steerable mobile information seeking system based on multi-LLM collaboration (Figure~\ref{fig:all-frame}). The multi-LLM architecture of DroidRetriever consists of three key modules: task decomposition, UI Copilot, and report synthesis. Users can express their information needs in natural language by voice or typed input, and in the task decomposition module, the system automatically filters candidate apps based on the user's description and breaks the task down into several sub-tasks to guide subsequent actions. The UI Copilot module executes these sub-tasks in sequence, automatically navigating to the target screens containing the desired information and capturing screenshots along the way, with a real-time progress dashboard that surfaces sub-task status and live exploration traces, where user can intervene and correct the system's navigation choices as needed at any time. Finally, in the report synthesis module, the system synthesize report by processing information extracted from app interfaces in accordance with user requirements. Key information is supplemented with citation links to facilitate content verification and referencing.


The main contributions of this work are summarized as follows:
\begin{itemize}
    \item We introduce DroidRetriever, a novel information integration system that utilizes mobile UI Copilot and collaborative multi-LLM, helping users more efficiently access daily in-app information.
    \item We designed a transparent and steerable user participation mechanism. The system provides a unified dashboard of task progress and execution traces. Users can check, interrupt, and take over at any time, and the system pauses on privacy sensitive or high risk screens with vibration and textual prompts.
    \item Across 35 tasks over 24 apps, experiments and user studies (51 participants in total) demonstrate improvements in coverage, accuracy, transparency, and reduced workload, with participants reporting  higher perceived transparency and a stronger sense of control.
\end{itemize}

\section{Background and Related Work}
Related work ranges from LLM-driven web search agents to mobile task-execution systems, concluding with approaches that support user intervention and takeover for controllable automation.

\subsection{LLM-Driven Web Search Agents}

In summarizing and presenting results, LLMs rank, compare, and consolidate information from diverse sources, adapting outputs to user needs. Open-source tools like Lepton~\cite{Lepton} and Perplexica~\cite{Perplexica} synthesize data from platforms such as Bing, Reddit, and YouTube, while Lumina~\cite{Lumina} employs LLMs to evaluate and refine search results for relevance. For domain-specific tasks, tools like Genspark~\cite{Genspark} and Wanderboat~\cite{Wanderboat} structure travel or product information into user-friendly formats, and Devv~\cite{Devv} organizes programming-related snippets for clarity.

However, most LLM-based agents rely on general search engines (e.g., Google) through APIs, limiting their ability to handle personalized or app-specific queries. Critical information-such as dynamic pricing in food delivery apps-often resides within individual apps and varies based on user-specific factors like location or discounts. Existing approaches fail to address such in-app information tasks unless explicitly supported by dedicated APIs.

\subsection{Mobile Agents for Information Seeking}

Existing tools, such as iOS's Siri or Honor's YoYo, assist users in completing specific information seeking tasks on mobile devices by interpreting natural language commands (e.g., checking the weather or managing schedules). These systems typically rely on intent recognition and pre-configured APIs to fetch answers directly, but their rigidity limits their ability to address diverse user needs. Pioneering work also explored dialog pause-and-correct and privacy approaches. SOVITE~\cite{Toby20-UIST} pioneered a multi-modal repair mechanism, enabling users to visually rectify dialog misunderstandings. Pinalite~\cite{Toby20-CSCW} addressed critical privacy barriers by establishing the capability for privacy-preserving sharing of GUI-based automation scripts, while Crepe~\cite{Crepe} democratizes mobile screen data collection by privacy-preserving graph query, providing both precise data localization and strong generalizability. Recent advancements in LLMs have enabled more dynamic task fulfillment, leveraging their natural language understanding and planning capabilities~\cite{Taskbench2023, Screenai2024, MM-Navigator, GPTVoiceTasker2024, EasyAsk2024, Nav-Nudge2024}. For instance, Wang~\cite{Wang2023} introduced a method using UI view hierarchies and LLMs to interpret screen content, while Autodroid~\cite{Autodroid2024} extended this by predicting step-by-step operations. To mitigate challenges with view hierarchies, visual-based approaches like VisionTasker~\cite{VisionTasker2024} and multi-modal LLMs (e.g., Ferret-UI~\cite{Ferret2024}, CogAgent~\cite{Cogagent2024}, Mobile-Agent~\cite{Mobile-agent2024}) have emerged, achieving near-human performance in task automation. However, these studies primarily focus on simple functional tasks (e.g., sending messages or setting alarms) and are not optimized for information retrieval. For example, when visiting multiple search results, existing UI agents are prone to repeatedly accessing the same result~\cite{Zhu25}.

Recent efforts to optimize mobile agents for multi-application tasks, such as Mobile-Agent-v2~\cite{Mobile-agent22024}, Mobile-Agent-v3~\cite{Mobile-Agent-v3}, MobileGPT~\cite{MobileGPT}, and AppAgent~\cite{AppAgent}, typically decompose tasks into sub-tasks and leverage both short and long term memory to manage progress and operational knowledge. However, their memory mechanisms often conflate action sequences with informational outcomes, which can result in missing critical results or introducing irrelevant ones. Moreover, most approaches rely solely on LLMs to extract and summarize textual fragments, without establishing one-to-one correspondence with the actual screen state or ensuring verifiable evidence tracing. In addition, they lack controllable stopping criteria and budgeted search strategies. They freely generate excessive exploration steps without parameterizable browsing limits or information-gain thresholds, making the scope of search difficult to constrain.

Systems that support user interruption and takeover, such as Claude Computer Use~\cite{Claude-Computer-use}, OpenAI Operator~\cite{OpenAI-Operator}, Agent S2~\cite{Agent-S2}, and VSA~\cite{VSA}, introduce mechanisms including pre-execution confirmation, dynamic interruption~\cite{Claude-Computer-use, Agent-S2}, and proactive pausing~\cite{OpenAI-Operator}. Yet, these systems typically present execution traces through dialog flows or video replays, without structured progress monitoring, explicit state markers, or indications of potential errors. As a result, users who wish to inspect or take over must reconstruct the entire history themselves, which imposes a considerable cognitive burden. These limitations highlight that achieving transparent and controllable information gathering on mobile devices remains an open challenge.

\section{Design Goal}
\label{Design Goal}

We propose that DroidRetriever, or an effective system designed to assist users in retrieving in-app information should be guided by design goals in the following three main aspects:



\textbf{(1) Automated planning and navigation}: Mobile information seeking is performed on small screens and the process is often lengthy and error-prone~\cite{Gupta16, Yu12, Jungselius25}. It is further hindered by cross-app "information islands", forcing users to shuttle, compare, integrate, and reconcile information across apps, escalating steps, time, and error risks~\cite{Hornbak24, Liang24, Kotut24, Karlson10}. 
Facing all these, DroidRetriever should not only interpret user intent from natural language commands or responses, but also: \textbf{[DG1A]} intelligently select the apps, \textbf{[DG1B]} decompose tasks into app-specific sub-tasks with a bounded search scope, \textbf{[DG1C]} devise navigation strategies to reach target screens and capture required information (e.g., via screenshots).

\textbf{(2) Summarization and result presentation}: On mobile devices, dispersed information and small screens hinder efficient comparison and consolidation. Structured aggregation and clear grouping can significantly reduce unnecessary revisits and comparisons, improving efficiency, coverage, and accuracy~\cite{Karlson06, Siu14, Chakraborty15, Arguello16, Mao18, Wang21}. Presenting evidence and its sources with the result, rather than providing abstract explanations or only relevance scores, better supports verification and fosters trust~\cite{Singh19, Juneja24}. Therefore, after navigation, DroidRetriever is designed to \textbf{[DG2A]} process extracted information from screenshots, \textbf{[DG2B]} synthesize structured reports in the form of tabular comparisons or narrative summaries, and \textbf{[DG2C]} automatically cite source references for key findings to enable user verification.



\textbf{(3) User intervention friendly}: Mobile agents performing multi-step tasks should enable transparent collaboration and steerable intervention. Related research offers guiding principles suggesting that action previews, user-controllable adjustments, and explanatory debugging help users maintain awareness and recover from errors~\cite{Horvitz1999, John04, Kulesza15, Amershi19, Wischnewski23}. Moreover, recent mobile-agent work recommends proactively pausing at decision points for user choices or assistance, especially in high-risk or privacy-sensitive cases~\cite{Qihang25, Yi25}. Accordingly, DroidRetriever is designed to: \textbf{[DG3A]} present clear and transparent interfaces that surface task progress and key intermediate results, enabling users to easily inspect and verify the agent’s ongoing actions; \textbf{[DG3B]} keep the agent steerable by allowing users to interrupt and take over at any time to correct potential deviations, reducing the need for constant supervision; and \textbf{[DG3C]} proactively pause and request confirmation for privacy and risk-sensitive operations (e.g., payments, addresses, or account changes).

\section{DroidRetriever}
Guided by the design goals, we developed DroidRetriever (Figure~\ref{fig:all-frame}, left), a multi-LLM system that assists users in efficiently retrieving in-app information.

\subsection{Example Usage Scenario}
Irene is at a shopping mall with her friend. During their conversation, the recently popular "Crumbl Cookies" comes up, and Irene gets a craving. She wants to check if there is a store nearby and find out the earliest she could pick up an order. However, she finds it inconvenient to manually open, navigate, and compare information across multiple apps while walking and chatting. Instead, she activates DroidRetriever with a simple voice command: "\textbf{Can you check the earliest pickup time for the nearest Crumbl?}"

Without any further action required from Irene, DroidRetriever identifies Uber Eats and DoorDash on her phone as the relevant applications and initiates an autonomous task. On the DroidRetriever dashboard, Irene sees the task broken down into clear, high-level steps: 1. Find "Crumbl" on DoorDash and 2. Find "Crumbl" on Uber Eats. Confident that the system is handling the search, she puts her phone away and goes to grab a coffee with her friend.

A short while later, her phone vibrates. DroidRetriever has paused its execution and sent a notification requesting user intervention. The dashboard indicates that the process was halted because the Uber Eats app required a login. Irene quickly completes the login process and then taps the "Back to Auto" button on the DroidRetriever interface, seamlessly handing control back to the system to complete its task.

Moments later, the process is finished. DroidRetriever presents a consolidated report at the bottom of its dashboard. The report compares the nearest stores found on both apps and highlights the fastest option: "Pickup available in 5 minutes via Uber Eats." To confirm the details, Irene taps the source citation link. This action reveals the evidential screenshot, with the exact pickup time and store address automatically highlighted, verifying the information's validity. With this reliable and quickly-obtained information, Irene can now effectively plan her time to pick up her cookies.


\subsection{Method Overview}
\label{Method Overview}
\begin{figure*}[]
    \captionsetup{justification=justified, singlelinecheck=false}
    \Description[The auto part of DroidRetriever.]{Figure 2: shows the auto part of DroidRetriever. It comprises three key modules: task decomposition, UI copilot, and report synthesis. The figure uses the query “seeking hotels near Universal Studios Japan for two people, September 12–14. How to choose between Expedia and Booking.com” as an example to illustrate how the three modules collaborate in practice.}
    \centering
    \includegraphics[width=0.93\linewidth]{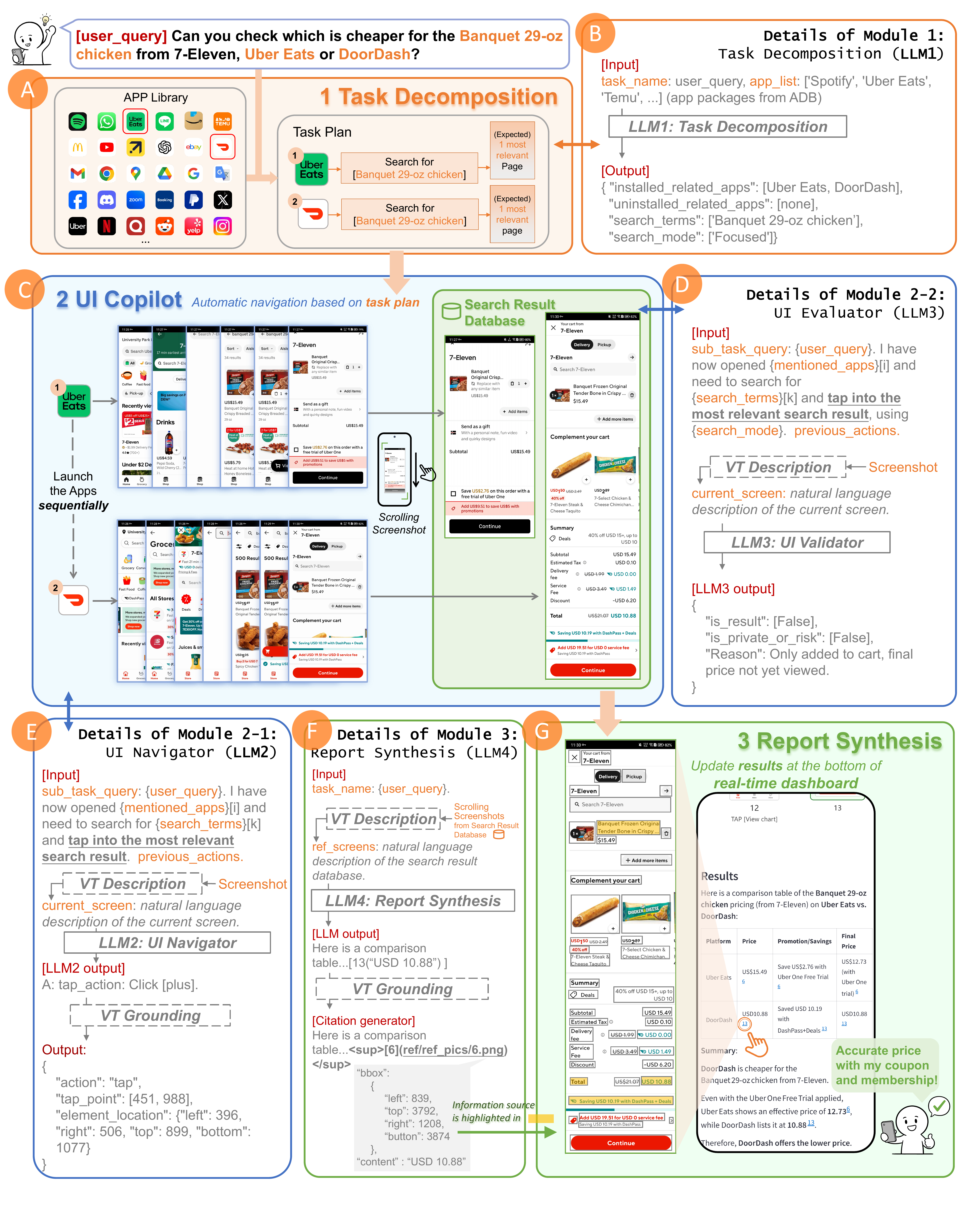}
    \caption{The Automated Workflow of DroidRetriever (excluding manual intervention). It includes 3 modules: task decomposition, UI copilot, and report synthesis. The system takes natural language commands from users, decomposes tasks, navigates automatically, and finally displays the report at the bottom of the dashboard.}
    \label{fig:framework}
    
\end{figure*}

DroidRetriever employs a multi-LLM architecture. Figure~\ref{fig:framework} illustrates the automated workflow without manual intervention, which employs a multi-LLM architecture with three core modules: task decomposition (DG1), UI copilot (DG1), and report synthesis (DG2). Upon receiving a natural language query, the task decomposition module breaks it into sub-tasks, identifies the most relevant installed apps, and assigns app-specific sub-tasks.

The UI copilot module sequentially opens the selected apps and autonomously navigates to target pages, with each step visually indicated via on-screen text prompts and highlighted UI elements to enhance transparency. During navigation, scrolling screenshots of all visited pages are captured and appended to the search results database. 

The report synthesis module processes the captured screenshots, extracts relevant data, and generates a clear and concise summary for the user. Key points are annotated with their source screen numbers and visually highlighted in the interface for quick verification (DG2).

DroidRetriever maintains an integrated dashboard (Figure~\ref{fig:intervention-interface}) that combines (i) sub-task progress for each app and search term, and (ii) real-time exploration maps showing the traversed pages and current position. This design allows users to monitor the process and intervene at any moment, eliminating the need for constant supervision (DG3, Figure~\ref{fig:intervention}).

\begin{figure*}[t]
    \centering
    \Description[Three search modes]{Figure 3: illustrates three search modes: focused, list-view, and multi-page. The left section shows the focused mode, which terminates after extracting target information from a single detailed page (e.g., restaurant menu details). The middle section depicts the list-view mode, designed for processing or comparing basic information across multiple items (e.g., flight options), halting at the overview page to avoid unnecessary detail navigation. The right section illustrates the multi-page mode, which sequentially navigates unvisited pages to gather detailed insights (e.g., product comparisons), with a user-defined maximum limit.}
    \includegraphics[width=0.85\linewidth]{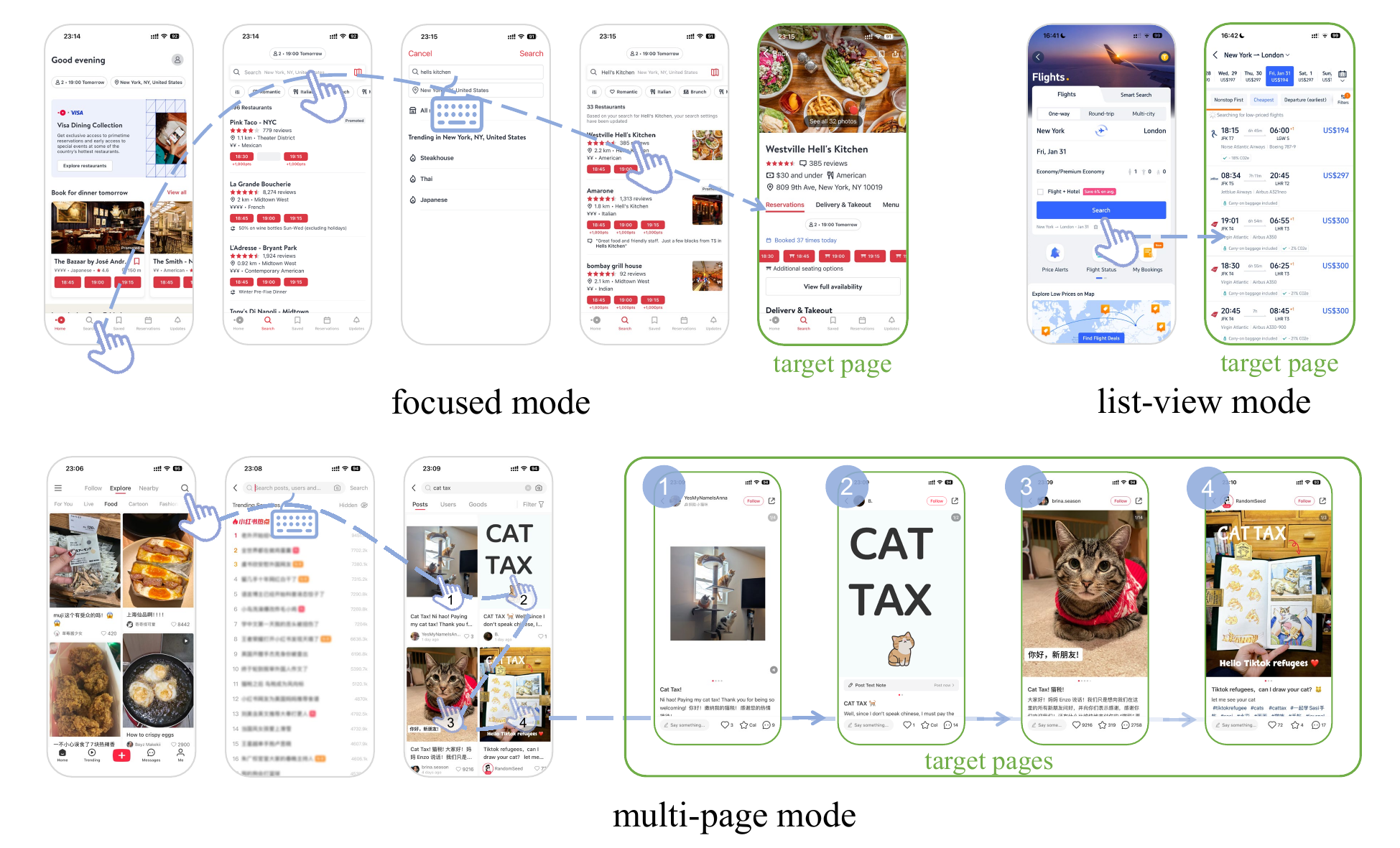} 
    \caption{Overview of Page-level decomposition, showing focused mode, list-view mode, and multi-page mode.}

    \label{fig:search_mode}
\end{figure*}

\subsection{Multi-LLM Framework}
\label{Multi-LLM Framework}

LLM-based mobile agents rely on system prompts to guide their behavior, but as conversations lengthen in complex tasks, they often conflate action traces with information results, leading to forgotten valid candidates, redundant exploration, and omissions or inconsistencies in the final report. Even with memory, most systems extract and summarize information by LLM without grounding to the actual screen state, limiting evidence traceability and reliable takeover in information retrieval tasks.


To address these limitations, we introduce a multi-LLM framework, DroidRetriever, that structures mobile information retrieval into three phases: (1) task decomposition (\textbf{\textit{LLM1}}) performs app-level, search-term-level, and page-level decomposition to derive executable paths (Figure~\ref{fig:framework}A-B) (DG1A-B); (2) UI copilot (\textbf{\textit{LLM2}}: UI Navigator; \textbf{\textit{LLM3}}: UI Evaluator) executes context-aware navigation and confirms arrival on the Search Results List Page or Result Details Page, pausing on privacy-sensitive operations or high-risk actions and capturing scrolling screenshots when a subtask completes (Figure~\ref{fig:framework}C-E) (DG1C); and (3) report synthesis (\textbf{\textit{LLM4}}) aggregates citation-linked snapshots into a Markdown report with tabular comparisons or narrative summaries (Figure~\ref{fig:framework}F-G) (DG2). This role specialization keeps scope controllable, separates navigation from information memory, and preserves transparent provenance.

\subsection{Task Decomposition}


We design a three-level decomposition to ensure robust cross-page retrieval while controlling interaction cost. The task decomposition module identifies candidate apps from the user’s query (with voice input transcribed by the SenseVoice-Small~\cite{SenseVoice}) and generates app-specific sub-task descriptions for execution. As shown in Figure~\ref{fig:framework} A-B, the LLM processes the natural language command and the installed app list (collected via ADB) to perform three key steps: (1) selecting relevant apps (app-level decomposition), (2) generating search terms (search-term-level decomposition), and (3) determining the optimal search modes (page-level decomposition)(DG1A-B).


The app-wise decomposition splits tasks into sub-tasks across applications by first identifying installed apps through package name parsing. The LLM prioritizes apps explicitly mentioned in the user's query for sub-task assignment; when unspecified or unavailable, it selects up to three relevant installed alternatives. These apps are then queued to be queried (Figure~\ref{fig:framework} A).


Search-terms-wise decomposition structures sub-tasks hierarchically within each application, where each sub-task corresponds to a specific app with a term. It first determines whether to use the app's search function or direct navigation. For search-based tasks, it produces multiple related terms to ensure comprehensive coverage; otherwise, it initiates direct UI Copilot to target screens. For queries that generate search terms, it rewrites the sub-task as "Open <APP>, search <Search term>, and tap into one search result"; otherwise, it keeps the original query.


After search, each sub-task potentially yields multiple results. We optimize three decomposition modes to balance information volume and task complexity (Figure~\ref{fig:search_mode}): (1) the \textit{\textbf{focused}} mode, which stops after extracting target information from a single Result Details Page (e.g., restaurant menu details); (2) the \textit{\textbf{list-view}} mode, for processing or comparing basic attributes across items (e.g., flight options) on the Search Results List Page, avoiding unnecessary detail navigation; and (3) the \textit{\textbf{multi-page}} mode, which sequentially visits unvisited Result Details Pages to gather deeper insights with a user-defined maximum. These modes minimize navigational errors and prevent LLM1 reporting inaccuracies caused by information overload.

\subsection{UI Copilot}
\label{UI Copilot}
The fully automated workflow of the UI Copilot consists of 3 components: the UI navigator (LLM2) for automatic execution, the UI evaluator (LLM3) for progress verification and privacy or risk checks, and the Search Result Database for storing target UI states (DG1C). User intervention during navigation is discussed in Section~\ref{Intervention During Navigation}.
 
The UI navigator (LLM2) automatically navigates to the target interface according to sub-tasks, as shown in Figure~\ref{fig:framework}C. Navigation involves two key steps: UI comprehension and action planning \& execution. For UI comprehension, we use VisionTasker (VT), an open-source vision-based framework~\cite{VisionTasker2024}; for action planning, LLM2 makes decisions and executes. 
VT provides two core capabilities: VT Description, which produces textual summaries of screenshots, and VT Grounding, which precisely identifies and locates UI elements. VT captures screenshots and applies three lightweight computer vision models-YOLOv8~\cite{Jocher_YOLO_by_Ultralytics_2023} trained on RICO~\cite{deka2017rico} for UI element detection, PaddleOCR\footnote{https://github.com/PaddlePaddle/PaddleOCR} for text extraction, and a CLIP model fine-tuned on IconSeer~\cite{li2023you} to interpret icons without textual context to analyze UI elements. This process yields a natural language description of the interface (VT Description) and the semantic details plus spatial coordinates of elements (VT Grounding). 
For action planning \& execution, LLM2 dynamically predicts single-step actions (tap, input, scroll, swipe, long press, open app, back) based on the current interface state and the operation history. After selecting the target element and action, VT Grounding converts it to screen coordinates, and the execution engine simulates touch events. The interface then updates-VT reanalyzes the new UI-and LLM2 iterates the next move until the subtask completes or a termination condition is met.


The UI Evaluator (LLM3) determines sub-task completion and stops the navigation, and flags privacy-sensitive operations or high-risk actions for pause and intervention. For tasks without generating search terms (focused mode by default), the system stops once it navigates to the interface with the most relevant information.

For in-app search tasks (in one of: focused, list-view, or multi-page mode), after entering a search term and tapping the search button, the system opens a "\textbf{Search Results List Page}" showing brief results, and taps a result will navigate to the "\textbf{Result Details Page}" for further information. In list-view mode, LLM3 confirms completion upon reaching the "search results list page," while in focused and multi-page modes, completion requires navigating to the "result details page." LLM3 determines sub-task completion by assessing whether the natural-language UI description contains sufficient information (Figure~\ref{fig:framework} D). The multi-page mode handles parallel sub-tasks (e.g., collecting smartphone reviews from multiple Quora answers): after each result, it returns to the "search results List page," selects the next unvisited result, and repeats until reaching the user-defined browsing limit, balancing time and completeness. To avoid duplicates, visited results are masked by editing VT description or adding cues in the prompt; if same pages are revisited twice, the system auto-scrolls to load new results.

Upon completing a sub-task (i.e., locating the target screen), scrolling screenshots are captured (Appendix~\ref{appendix:scrolling-screenshot} Figure~\ref{fig:scrollingscrshot}) by performing four downward slides (each covering about 2/3 of the screen) and stitching the captures into a long-page screenshot. This approach effectively captures key content, which typically appears within the first two screens according to UI design principles~\cite{robbins2007user, lagun2014towards}. The system stores the screenshot in the Search Result Database for reporting, then checks for pending sub-tasks in page-wise, search-term-wise, and app-wise order. If incomplete sub-tasks exist, it returns to the appropriate branching point (e.g., search results page or app home screen) to proceed by pressing Back, force-reopening the app, or closing to launch the next app; otherwise, it proceeds to report synthesis.

\begin{figure*}[]
    \captionsetup{justification=justified, singlelinecheck=false}
    \Description[The workflow of Intervention.]{Figure 4: illustrates the intervention and feedback workflow. Users first check questions about "Boba Tea" on Quora during task decomposition to identify key aspects of focus. They then navigate the interface by opening the Quora app, entering search keywords, and accessing the search result page. During intervention (Intervention^a), users can directly control actions such as tapping or inputting to correct errors; after manual adjustments, clicking the "Return to Auto" button resumes automation. Users can also take screenshots at any time via intervention (Intervention^b) and save the current interface to the search result database for comprehensive information capture. Additionally, users can decide whether to terminate the task and generate a report through intervention (Intervention^c).}
    \centering
    \includegraphics[width=1\linewidth]{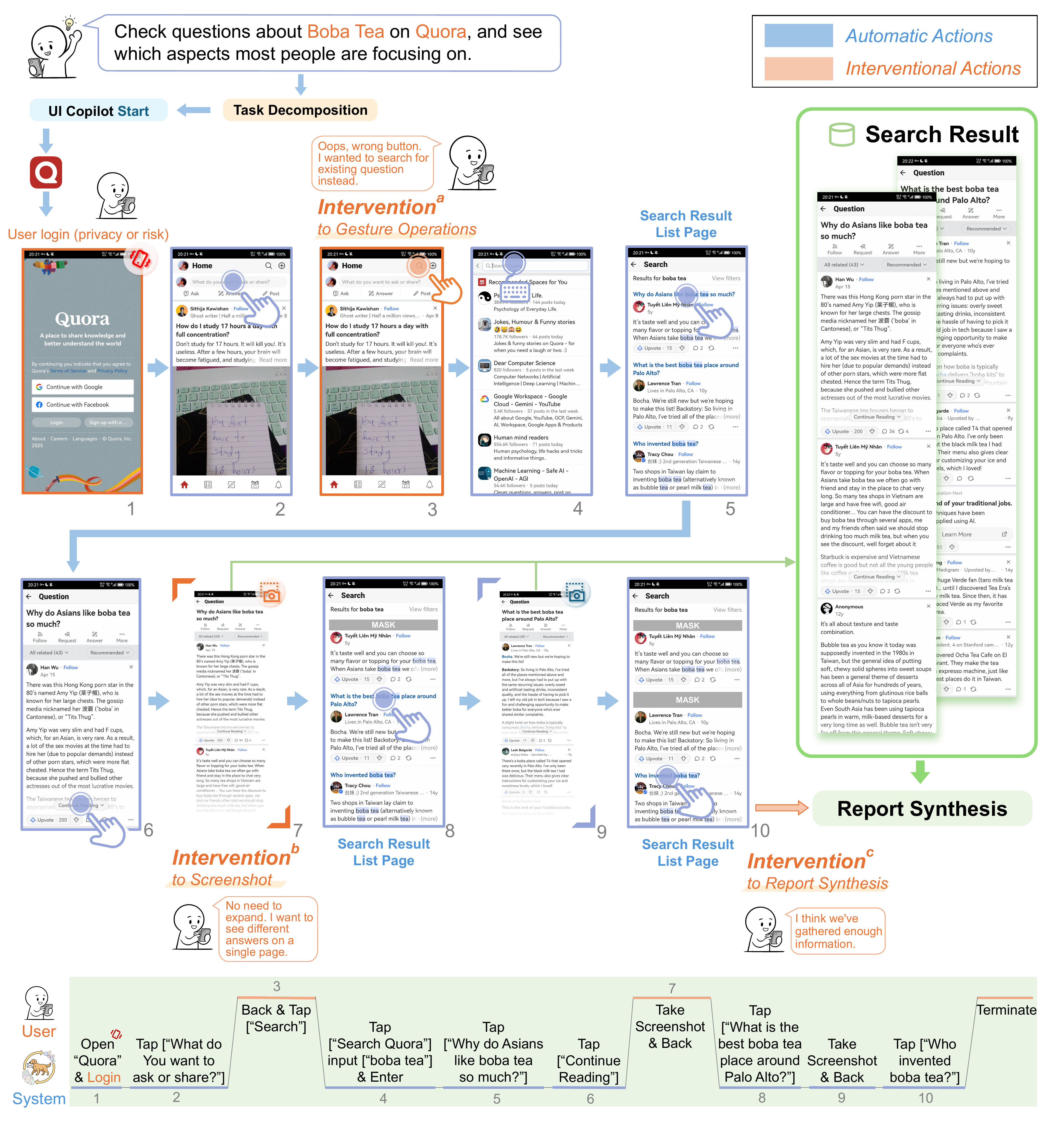}
    \caption{Intervention mechanisms: $Intervention^a$ requires gesture operations during intervention, such as tapping and text input, and also includes intervention for proactive alerts on privacy-sensitive operations and high-risk actions. $Intervention^b$ lets the user take a screenshot and save the current interface to the search results database. $Intervention^c$ signifies the intention to terminate the UI copilot.}
    \label{fig:intervention}
\end{figure*}

During the whole navigation process, our system previews each LLM-planned operation to users via text notifications and highlighted overlays before execution (Figure~\ref{fig:intervention}(b)), enabling user intervention and corrections when necessary. Once the UI Evaluator (LLM3) detects a privacy-sensitive or high-risk interface (e.g., personal information, payments, file deletion), the system pauses, triggers vibration and a notification to prompt user to take over.

\begin{figure*}[]
    \captionsetup{justification=justified, singlelinecheck=false}
    \Description[The 2 interfaces for Intervention.]{Figure 5: shows the widget for user to taker over on the phone. During intervention (Intervention^a), users can directly control actions such as tapping or inputting to correct errors; after manual adjustments, clicking the "Return to Auto" button resumes automation. Users can also take screenshots at any time via intervention (Intervention^b) and save the current interface to the search result database for comprehensive information capture. Additionally, users can decide whether to terminate the task and generate a report through intervention (Intervention^c). These three basic intervention buttons are folded into the expansion button. (c) presents the real-time overview of sub-task progress and the execution trace. The user can leave it run, inspect at any time, and interrupt if needed.}
    \centering
    \includegraphics[width=1\linewidth]{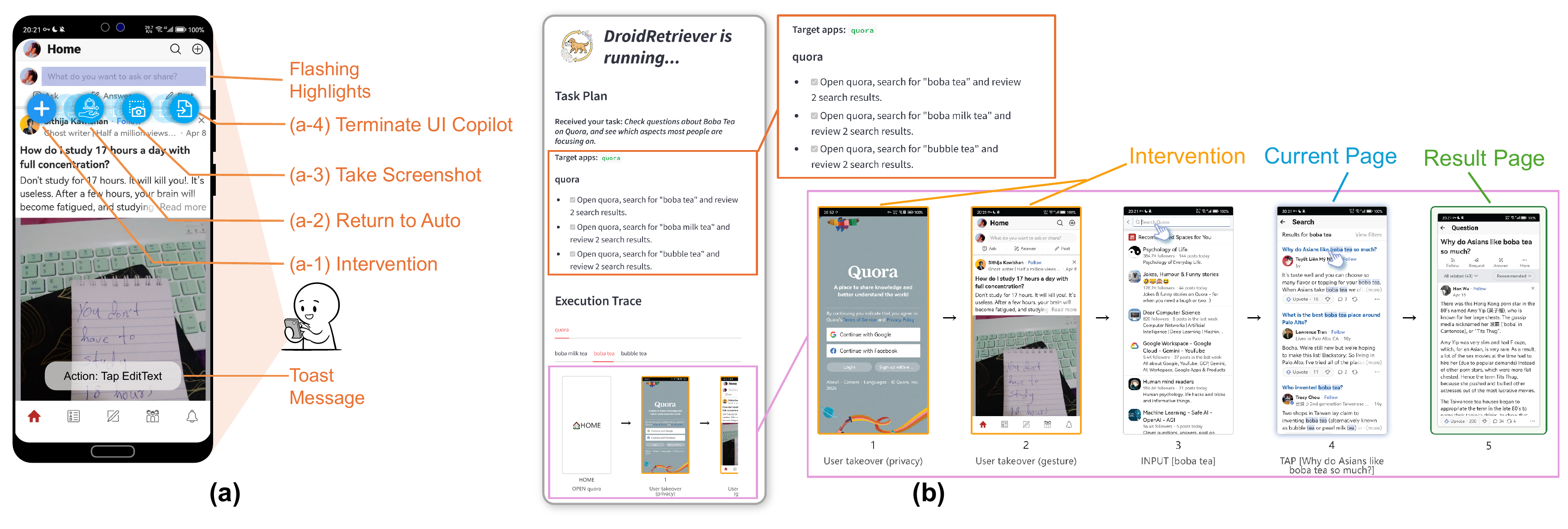}
    \caption{User interfaces: (a) shows the intervention widget: (a-1) intervention - interrupt and take over, (a-2) tap to return to automatic mode, (a-3) take a screenshot and save to the search results database, and (a-4) terminate the UI Copilot. (b) presents the real-time dashboard of sub-task progress and the execution trace. The user can leave it run, inspect at any time, and interrupt if needed.}
    \label{fig:intervention-interface}
\end{figure*}

\subsection{Intervention During Navigation}
\label{Intervention During Navigation}

As shown in Figure~\ref{fig:intervention-interface}(a), DroidRetriever features a lightweight intervention widget to enable human intervention: (a-1) intervene to interrupt and take over, (a-2) tap to resume automatic mode, (a-3) take a screenshot and save it to the search results database, and (a-4) terminate the UI copilot (DG3B). Buttons (a-2)-(a-4) are nested within (a-1) and hidden by default; tapping (a-1) reveals them. The widget is draggable to avoid obstructing content. During UI copilot, the predicted target element is highlighted by a semi-transparent purple rectangle that flashes three times, and a toast at the bottom provides specific instructions (e.g., "Tap [Texas]" or "Enter [McDonald] in the [Search] field").


\textbf{Intervention\textsuperscript{a}} and "Intervention" button in Figure~\ref{fig:intervention-interface}(a-1) enables users to take over with gesture operations in an intuitive way, whether triggered by proactive pauses on privacy-sensitive or high-risk pages, or initiated from the progress dashboard when issues are spotted (Figure~\ref{fig:intervention}, Figure~\ref{fig:intervention-interface}(b)). After manual adjustments, tapping "Return to Auto" (Figure~\ref{fig:intervention-interface}(a-2)) resumes automation. During intervention, the system neither records nor interferes with user actions, protecting privacy, especially for sensitive inputs such as passwords. The system logs the intervention with a fixed marker in the execution trace, reassesses task progress from the updated UI state, and reflects on the intervention to recover the correct path.


\textbf{Intervention\textsuperscript{b}} enables users to take screenshots at any time to save the current interface to the search database (Figure~\ref{fig:intervention-interface}(b-3)) for the final report, as illustrated in Figure~\ref{fig:intervention}. These user-initiated screenshots are automatically processed using the same scrolling capture mechanism as system-generated screenshots, ensuring comprehensive information capture. All screenshots, whether system-generated or user-captured, are stored in a unified search database with unique IDs. With \textbf{intervention\textsuperscript{c}} in Figure~\ref{fig:intervention}, users can determine whether to terminate UI Copilot at any moment (Figure~\ref{fig:intervention-interface}(a-4)) and generate the report based on the search database up to that moment.


Real-time dashboard of sub-task progress and the execution trace (Figure~\ref{fig:intervention-interface}(b)) let users leave the system running and check progress with low effort, compared with long dialog logs or video replays (DG3A). The task decomposition panel lists all sub-tasks and marks them as completed upon finish. The execution trace is organized by app and search term in tabs. Screenshots with an orange border indicate user intervention, blue highlights show the current interface, and green marks milestone pages saved to the search results database. The final report will be displayed at the bottom of this view. The interface was implemented with Streamlit, which continuously visualizes the runtime logs generated by DroidRetriever. During execution, the system records detailed logs, including task decomposition, execution screenshots, click coordinates (rendered with a fingertip icon at the corresponding position), navigation flow, intervention events, and progress status. Streamlit automatically refreshes and reads the latest log files every two seconds to display the system’s operations in real time.

\subsection{Report Synthesis}
\label{Report Synthesis}

The Report Synthesis (LLM4) processes natural language descriptions (generated by the VT description module from search result screenshots) to schematize and refine information into coherent reports tailored to user needs, enhancing readability and filtering irrelevant content (DG2). Depending on the task, it generates reports in two formats: (1) tabular comparisons (e.g., for product evaluations), structured by dimensions like price, features, and performance; or (2) narrative summaries (for general tasks), integrating insights from multiple screenshots into a clear, structured format with highlighted key points.


We preserve original screenshots and leverage VT Grounding to make reports auditable end-to-end and ensure transparency. The report cites key points (Figure~\ref{fig:framework} G) by linking them to source screenshots. The VT Grounding module processes scrolling screenshots from the search database, first segmenting them (with white padding if needed) to match the screen height of VT's training set (RICO dataset~\cite{deka2017rico}), ensuring compatibility with the trained object detection model. Each segment is analyzed to locate UI elements, with bounding box offsets calculated to map their true positions in the original scrolling screenshot for visually highlighting. The module then fuzzy-matches (threshold: 0.8) report content to these elements, mitigating LLM hallucinations by selecting the most similar reference. Citations are embedded via Markdown, with highlighted screenshot regions (Figure~\ref{fig:framework} G, right panel) and support for rich text (tables, links, emphasis).

\section{Study 1: Information Extraction Evaluation}
\label{sec:study-1}

In this section, we evaluate DroidRetriever’s ability to synthesize task-oriented reports from screenshots through quantitative and qualitative analysis. Specifically, we assess whether our method produces reports that cover the key information required by the task, present it clearly, and are easy for users to understand.

We particularly emphasize the report synthesis module, as it addresses a standalone and broadly applicable user need: aggregating content selected by the user across multiple screens and and clearly attribute the sources of information. For instance, when choosing a hotel, users often bounce between pages to compare ratings, locations, and facilities, taking notes to capture the key details. Report synthesis streamlines this process by consolidating dispersed information into a clear and well-supported comparison.

\subsection{Method}
\subsubsection{Procedure}

We conducted a controlled user study comparing DroidRetriever against human participants on 13 common information tasks spanning multiple domains (e.g., payments, maps, shopping, news, and social media). Both were presented with screenshots and required to extract key information, simulating typical mobile app interactions where users read, comprehend, and record data for decision or discussions. The complete list of tasks (translated into English) is presented in Table~\ref{tab:study1-task}. Task design incorporated varying levels of information complexity, with 54\% being simple tasks (less than 10 key points) and 46\% complex tasks (extensive information). Four core information processing capabilities were evaluated: summarization (three tasks involving condensing key text), comparison (two tasks requiring multi-dimensional tabular comparisons), processing (five tasks including sorting, filtering, calculating, and integrating data from multiple sources), and localization (three tasks focusing on multi-language interpretation and domain-specific explanation). All tasks used real scrolling screenshots from popular apps (>50M downloads on Google Play/Huawei AppGallery), mixing task-relevant information and distractions.

\begin{table*}[t]
    \centering
    \footnotesize
     \Description{Table 1: presents the 13 tasks in Study 1 along with their descriptions and functional categories. Summarization tasks include: listing Alipay services with password-free or auto-pay enabled, summarizing how to use the GTD work method on Zhihu, and checking the ticket refund policy on 12306. Comparison tasks include: comparing the specifications of OPPO Find X7 Ultra vs. VIVO X100 Ultra and comparing Xiaomi 14 256GB prices and deals on Taobao vs JD. Processing tasks include: showing taxi trips from May to August costing 15-20 yuan, calculating phone credit from October to August with a monthly average, identifying the top 3 most frequent movies across ranking charts, listing available afternoon trains from A to B on September 12, and listing delivered packages by express station. Localization tasks include: explaining the functions of the nutrients in a baby formula, describing how to disable private messages on Quora, and translating Red Velvet's latest post.}
    \caption{An Overview of Study 1: 13 Tasks, Capabilities, and Scoring Points.}
    {
    \begin{tabular}{c p{7cm} c c}
    \toprule
    \# & Task & Capabilities & Scoring Points (Count) \\
    \hline \hline
    1 & List Alipay services with password-free or auto-pay enabled. & Summarization & 16 \\ \hline
    2 & Summarize how to use the GTD work method on Zhihu. & Summarization & 10 \\ \hline
    3 & Check the ticket refund policy on 12306. & Summarization & 8 \\ \hline
    4 & Compare OPPO Find X7 Ultra vs. VIVO X100 Ultra specs. & Comparison & 12 \\ \hline
    5 & Compare Xiaomi 14 256GB prices and deals on Taobao vs JD. & Comparison & 6 \\ \hline
    6 & List available afternoon trains from A to B on Sept 12. & Processing & 20 \\ \hline
    7 & Show me taxi trips from May to August costing 15-20 yuan. & Processing & 12 \\ \hline
    8 & List delivered packages by express station. & Processing & 3 \\ \hline
    9 & Identify top 3 most frequent movies across ranking charts. & Processing & 3 \\ \hline
    10 & Calculate phone credit from Oct to Aug with monthly average. & Processing & 2 \\ \hline
    11 & Explain the functions of the nutrients in this baby formula. & Localization & 10 \\ \hline
    12 & Translate the technical post and discussions below into English. & Localization & 4 \\ \hline
    13 & How to disable private messages on Quora? & Localization & 2 \\ \hline
    \bottomrule
    \end{tabular}
    }
\label{tab:study1-task}
\end{table*}

\begin{figure*}
    \captionsetup{justification=justified, singlelinecheck=false}
    \Description{Figure 6: is the results of Study 1. Panel (a): coverage, accuracy, and redundancy rates comparing manual versus system-generated reports. Panel (b): overall quality ratings. A downward arrow (↓) indicates that lower values are better.  *** indicates a significant difference with p < .001, while ** indicates significance with p < .01.}
    \centering
    \includegraphics[width=0.8\linewidth]{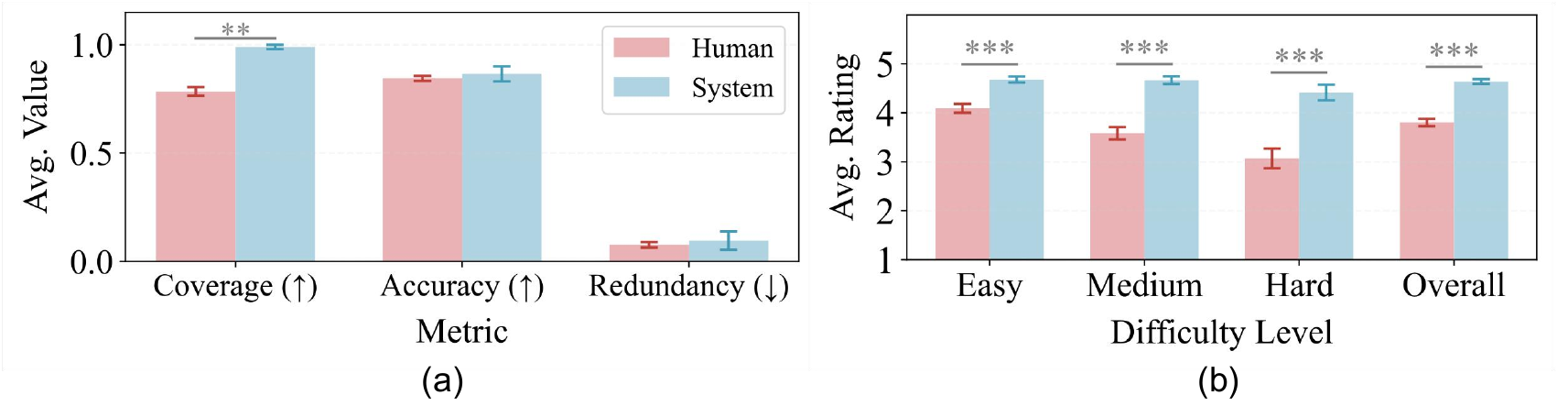}
    \caption{\textcolor{black}{Results of Study 1: (a) Coverage, accuracy, and redundancy rates for manual vs. system-generated reports; (b) Overall quality ratings. $\downarrow$ indicates lower is better. *** indicates a significant difference with \( p < .001 \), while ** indicates significance with \( p < .01 \).}}
    \label{fig:study1}
\end{figure*}


For DroidRetriever, we only utilized the report synthesis module to automatically generate reports from screenshots, contrasting with the manual process performed by human participants. For confidentiality reasons, we used a fixed set of static screenshots as input, captured from real-world mobile application interfaces and selected to be highly relevant to the task goals and to cover representative, challenging scenarios. Each participant was assigned on a desktop platform and explicitly instructed: "Please collect, as completely and accurately as possible, all information you believe is useful for accomplishing the specified task." The platform simulated a mobile scrolling experience, presenting a sequence of interface screenshots (with vertical scrolling and "Previous"/"Next" navigation). Participants could freely excerpt required information in a "Notes" area. Participants were not required to write a formal report, and could record in any format to reflect real-life "capture-and-notes" habits. The platform supported copying any content from the screenshots and allowed the use of any auxiliary tools except LLM-based content generation (e.g., search engines, calculators, translation services).

We recruited 16 participants (aged 21-34, 3 females) from a local university, selecting them for their extensive smartphone experience (daily usage >3 hours), and over 6 months of LLM application experience (e.g., ChatGPT, Copilot). After training on task requirements and platform operation, participants completed 13 tasks and documented information in a text box. Post-task, they rated difficulty (5-point Likert scale) and evaluated their own reports against DroidRetriever's. The study concluded with a 5-minute semi-structured interview on their DroidRetriever experience. The 60-minute session (on average) compensated participants at \$6 (USD, slightly above the local average hourly wage.)

\subsubsection{Metrics}


For each task, two authors independently identified the minimum essential \textbf{Scoring Points} required to accurately and completely answer the information task, then reached consensus (the details of scoring points are listed in Appendix~\ref{appendix:study1-scoring-points}). The number of scoring points is summarized in Table~\ref{tab:study1-task}: 31\% of tasks have more than 10 points, 31\% have 5-10 points, and 38\% have 1-5 points. These scoring points served as the basis for metric computation. We reported the mean values across all participants and tasks. The metrics include:

\begin{itemize}[leftmargin=*, itemindent=0pt]
    \item\textbf{Coverage}: Proportion of predefined scoring points mentioned in the report (regardless of accuracy) to the total required points. Higher values indicate greater comprehensiveness.
    \item\textbf{Accuracy}: Proportion of correctly documented scoring points to total documented scoring points in the report, which measures reliability of extracted information. Omitted points of the reports are not factored into this metric.
    \item\textbf{Redundancy}: Ratio of irrelevant points to total documented points. Lower redundancy indicate more concise reporting.
\end{itemize}

Furthermore, we include the following subjective metrics:
\begin{itemize}[leftmargin=*, itemindent=0pt]
    \item\textbf{Overall Quality}: Participants rated both manual reports and DroidRetriever's outputs on three dimensions (accuracy, coverage, readability) using a \textcolor{black}{1}-5 scale. We calculated an overall quality score by averaging these ratings (see Appendix~\ref{appendix:study1-questionnaire} for questionnaire details).    
    \item\textbf{Task Difficulty}: Participants assessed each task difficulty using a three-level scale (simple, moderate, difficult).
\end{itemize}

\subsection{Results and Analysis}
\textcolor{black}{We used pairwise $t$-tests for Coverage, Accuracy, and Redundancy, and the Wilcoxon Signed-Rank Test for Overall Quality.} Figure~\ref{fig:study1}(a) demonstrates that our system achieves better coverage ($p=.0096$) while maintaining comparable accuracy ($p=.77$). System reports exhibited errors primarily in detail-oriented listing (Task 1) and mathematical operations (Task 9), reflecting LLM limitations regarding hallucinations and calculation. Despite these issues related to accuracy, the system produced more complete reports, covering a broader range of relevant information. In contrast, manual reporting showed information omissions primarily in complex tasks (5 and 11) involving technical terminology, while factual errors occurred in tasks requiring translation and calculations (9, 10, 13). These contrasting patterns align with our metric definitions: accuracy only considers the factual correctness of mentioned information, while coverage captures the completeness of the report.


Among 208 tasks, participants rated 119 (57\%) as easy, 61 (29\%) as moderate, and 28 (14\%) as difficult. Figure~\ref{fig:study1}(b) compares the overall quality ratings between manual and system-generated reports across all task instances ($N=208$). Our system-generated reports achieved a significantly higher mean rating ($p<.001$). These automated reports received higher ratings in 110 instances (53\%), equal ratings in 77 instances (37\%), and lower ratings in only 21 instances (10\%). The performance advantage was most notable in moderate and difficult cases, with 18 instances showing 3 points or more rating differences favoring automated reports. Participants attributed their lower manual ratings primarily to omitted or incomplete information (9 instances), calculation or interpretation errors (6), unclear expression or structure (4), and domain knowledge limitations (4). The 3 instances where manual reports scored higher both involved calculation errors of our system.


Participant interviews highlighted both strengths and aspects that could benefit from further enhancement. The most valued aspects included thorough content coverage (5 participants), well-structured reports (4), and precise information referencing (3), with one participant noting "even when errors occurred, I could quickly locate information in the original screenshot." However, limitations emerged regarding computational accuracy (5) and information redundancy (3). Suggested improvements included enhancing image reference methods with direct links to highlighted content (2) and increasing output conciseness (2). Participants also highlighted potential applications, such as deep analysis of figures, charts, and consumption data (6).

\section{Study 2: Usability Evaluation}

In this section, we further evaluate the capabilities of task decomposition and UI copilot modules. We also conducted a comparative study assessing our system against several intelligent information retrieval tools, including LLM-driven search engines (Qwen and ChatGPT), Claude Computer Use, and Mobile-Agent-v2. We focus on three questions: (1) Can task decomposition module correctly break down complex tasks into executable sub-tasks? (2) Can the system efficiently and accurately reach relevant interfaces and capture screenshots without user intervention? (3) How effective is our system compared to existing solutions in supporting information retrieval across single-App and multi-App tasks?

\subsection{Method}
\subsubsection{Procedure}
\label{study2-Procedure}

\begin{table*}[t]
    \centering
    \footnotesize
    \renewcommand{\arraystretch}{1.1}
    \Description{Table 2: (Single-App Tasks) summarizes 14 tasks in Study 2 with brief descriptions and the primary functional capability each task targets. Summarization tasks include: checking which permissions have been authorized to Meituan, listing community guidelines on Bilibili, listing ticket redemption rights for 12306 members, checking the driving time to Shanghai on Amap, and summarizing reviews of “Black Myth” on Zhihu and Marshall Middleton on Rednote. Processing tasks include: counting how many Eason songs are saved on QQ Music, finding meetings scheduled by Irene, checking current tuition and fees on Mobile Campus, and showing total ride expenses on Amap for last month. Localization tasks include: checking the features of the latest added monitor in a JD cart, retrieving RAM specifications from an item in a Taobao cart, listing default currency settlement units on SHEIN, and translating a Red Velvet notification on Weverse.}
    \caption{An Overview of Study 2: Single-App Tasks, Capabilities, Access Type and Scoring Points}
    {
    \begin{tabular}{c p{7.5cm} c c c}
        \hline
        \textbf{\#} & \textbf{Task} & \textbf{Capabilities} & \textbf{Access Type} & \textbf{Scoring Points} \\
        \hline \hline
        1 & Check which permissions have been authorized to Meituan. & Summarization & Login-Gated & 9 \\
        \hline
        2 & List community guidelines on Bilibili. & Summarization & Open-Web & 6 \\
        \hline
        3 & List ticket redemption rights for 12306 members. & Summarization & Open-Web & 4 \\
        \hline
        4 & Check the driving time to Shanghai using Amap. & Summarization & Open-Web & 1 \\
        \hline
        5 & Find and summarize reviews of ``Black Myth'' on Zhihu. & Summarization & Open-Web & 3 \\
        \hline
        6 & Find and summarize reviews of Marshall Middleton on Rednote. & Summarization & Open-Web & 3 \\
        \hline
        7 & Find out how many Eason songs I've saved on QQ Music. & Processing & Login-Gated & 5 \\
        \hline
        8 & Find meetings scheduled by Irene. & Processing & Login-Gated & 3 \\
        \hline
        9 & What are my current tuition and fees on Mobile Campus? & Processing & Login-Gated & 1 \\
        \hline
        10 & Show me my total ride expenses on Amap for last month. & Processing & Login-Gated & 1 \\
        \hline
        11 & Check the features of the latest added monitor in my JD cart. & Localization & Login-Gated & 10 \\
        \hline
        12 & Tell me the specifications of the RAM in my Taobao cart? & Localization & Login-Gated & 5 \\
        \hline
        13 & List all default currency settlement units on SHEIN. & Localization & Open-Web & 4 \\
        \hline
        14 & Translate Red Velvet notification from Weverse to Chinese. & Localization & Open-Web & 2 \\
        \hline
    \end{tabular}
    }
\label{tab:study2-single-app-task}
\end{table*}

\begin{table*}[t]
    \centering
    \footnotesize
    \renewcommand{\arraystretch}{1.1}
    \Description{Table 3: (Multi-App Tasks) lists 8 tasks in Study 2 that require coordinating information across multiple apps, with their targeted capabilities. Summarization tasks include: identifying popular songs by King Gnu, checking Orlando Disney travel guides on Ctrip, Zhihu, and Xiaohongshu, gauging fans’ sentiment about potential new Mission: Impossible films, and checking the latest emails in Gmail and NetEase Mail. Comparison tasks include: deciding whether to order a Big Mac from McDonald’s, Meituan, or Ele.me, and advising on purchasing a VIVO X100 Ultra from Taobao versus JD. Processing tasks include: checking available showtimes two days from now for “F1: The Movie.” Localization tasks include: retrieving liked playlists on Apple Music, Spotify, and QQ Music and suggesting likely preferred music styles with brief rationale.}
    \caption{An Overview of Study 2: Multi-App Tasks, Capabilities, Access Type and Scoring Points}
    {
    \begin{tabular}{c p{8cm} c c c}
        \hline
        \textbf{\#} & \textbf{Task} & \textbf{Capabilities} & \textbf{Access Type} & \textbf{Scoring Points} \\
        \hline \hline
        1 & What are some popular songs by King Gnu in Spotify? & Summarization & Open-Web & 6 \\
        \hline
        2 & Check Orlando Disney travel guides on Ctrip, Zhihu, and Rednote. & Summarization & Open-Web & 8 \\
        \hline
        3 & Find out how fans feel about the possibility of more \textit{Mission: Impossible} films. & Summarization & Open-Web & 6 \\
        \hline
        4 & Check my latest email in Gmail and NetEase Mail respectively. & Summarization & Login-Gated & 2 \\
        \hline
        5 & Ordering a Big Mac from McDonald's, Meituan or Ele.me? & Comparison & Open-Web & 10 \\
        \hline
        6 & Advise on purchasing VIVO X100 Ultra from Taobao or JD. & Comparison & Open-Web & 8 \\
        \hline
        7 & Check showtimes available the day after tomorrow of the movie \textit{F1: The Movie}. & Processing & Open-Web & 2 \\
        \hline
        8 & Check my liked playlists on Apple Music, Spotify, and QQ Music, and suggest music styles I’d probably like with some background info. & Localization & Login-Gated & 5 \\
        \hline
    \end{tabular}
    }
\label{tab:study2-multi-app-task}
\end{table*}

We constructed a set of 22 representative tasks, which covered single-App tasks (14 in total) and multi-App tasks (8 in total). The details are provided in Table~\ref{tab:study2-single-app-task} and Table~\ref{tab:study2-multi-app-task}. 

Similar to Study 1, this study employed tasks from widely-used applications (50M+ installations) spanning common domains: payment, map, lifestyle, e-commerce, news, and social media. The tasks were carefully designed over various levels of complexity along two dimensions: information volume (50\% with fewer than five data points versus 50\% with five or more) and procedural steps (41\% requiring five navigation steps or fewer, 32\% needing six to ten steps, and 27\% involving over ten steps). We employed Qwen-2.5-72B for LLM1 through LLM4 implementations.


\textbf{Task Decomposition Evaluation}: We conducted a systematic evaluation of the task decomposition module across 22 tasks. The evaluation first examined the module’s capability in app decomposition: whether it could correctly identify and select the most appropriate application(s) to accomplish a given task based on its description. We then examined its performance in search mode selection, analyzing classification accuracy across focused mode tasks (45\%), list-view mode tasks (36\%), and multi-page mode tasks (18\%). 


\textbf{UI Copilot Evaluation}: We further evaluated the UI Copilot module across 22 tasks, focusing on its accuracy in autonomous interface navigation and result capture. Our evaluation focused on whether the module could locate task-relevant interfaces and store them in the search result database. Task criteria were differentiated by search mode: for focused mode tasks, the module needed to correctly navigate to and capture a detailed page that satisfied all specified query conditions; for list-view mode tasks, it had to identify and store a results list page that contained multiple task-relevant items; and for multi-page mode tasks, the module was required to visit multiple relevant search result detail pages and capture scrolling screenshots.

\textbf{Comparison with Other Tools}: The study compared 6 conditions across both single-App tasks and multi-App tasks: (1) manual completion by participants, (2) using the DroidRetriever system, (3) using LLM-driven search engines by \textit{Qwen} (a widely used local solution), (4) using commercial LLM-driven search engines by \textit{ChatGPT} (a globally dominant solution), (5) using Claude Computer Use (an agent optimized for user takeover and compatible with mobile platforms), and (6) using Mobile-Agent-v2 (an agent specifically optimized for multi-APP task execution).

The study was conducted on a HUAWEI P20 smartphone (Android, 5.8-inch, 2244×1080). For our system, we employed Qwen-2.5-72B with local deployment for LLM1 through LLM4 implementations. For LLM-driven search engines, we used (1) the closed-source \textit{Qwen-2.5-2024-0919 model with its official Quark Search plug-in, and (2) \textit{ChatGPT-5.1} with web search enabled.} We adapted Claude Computer Use (Claude-Sonnet-3.5) which originally designed for desktops but mobile-compatible officially by mirroring the smartphone display to a desktop screen (while maintaining Claude's native 1366×768 input resolution) for clearer UI visibility. This hybrid setup enabled keyboard, mouse control and touch input, with Claude supporting two interventions: manual adjustment or command refinement by natural language. For Mobile-Agent-v2, we strictly followed its official implementation.

We recruited 22 participants (3 female, 13 male, aged 21-34) with demonstrated smartphone proficiency (>3 hours daily usage) from a local university. Participants received a 5-10 minute tutorial and hands-on practice with all the systems before the study. We adopted a cyclic counterbalancing design, where the \textcolor{black}{six} methods were rotated across tasks and participants to mitigate the learning effect. Participants were free to check and intervene at any time using their preferred methods to ensure the quality of the report. After each task, they rated their mental workload, certainty, and confidence in completing the task. Each participant completed the study in 90 minutes, compensated at \$8.5 (USD).

\subsubsection{Metrics}

For Task Decomposition Evaluation, we calculated Decomposition Accuracy at both the app and page levels, and for UI Copilot Evaluation, we assessed Navigation Accuracy, as described in Section~\ref{study2-Procedure}. For the Comparison with Other Tools evaluation, we designed the following metrics.

We used three metrics regarding cost: Time, Steps, and Token Count. Time measures the total duration to complete the task; for tasks requiring user intervention for our system, we also logged the completion time and the duration of user interaction; Steps records operational steps to identify redundancy; and Token Count estimates API cost during execution.

We also analyzed two metrics related to user intervention in our system. 
The Task-wise Intervention Rate measures the proportion of tasks in which participants intervened, while the Step-wise Intervention Rate measures the average proportion of steps that involved user intervention out of all steps performed for each task.

Report quality was assessed using the same metrics from Study 1: Coverage, Accuracy, and Redundancy. These metrics were calculated by comparing predefined scoring criteria with the information points in the reports (the details of scoring points are listed in Appendix~\ref{appendix:study2-scoring-points}).

To further explore how well these tools support completing information tasks, we introduced three self-reported metrics: Workload, Certainty, and Confidence, each rated on a 5-point Likert scale. Workload measured perceived mental effort, Certainty assessed the perceived clarity regarding the available information involved in completing the task, and Confidence evaluated the level of trust users have in their outcomes. Participants completed these ratings after all tasks. The questionnaire is provided in Appendix~\ref{appendix:study2-questionnaire}

\subsection{Results and Analysis}
Overall, the app-level decomposition accuracy across 22 tasks was 96\%. Only one case (Task 7, a multi-app task) included an irrelevant app. In practice, this had no impact on outcomes: the additional app provided no useful information during execution and was therefore excluded in the final report.

\begin{figure*}[]
    \captionsetup{justification=justified, singlelinecheck=false}
    \Description[Study 2]{Figure 7: is the results of Study 2, including task decomposition and a comparative evaluation of Human, DroidRetriever, LLM-driven search engines (Qwen \& ChatGPT), Claude Computer Use, and Mobile-Agent-v2. (a) Page-level decomposition confusion matrix. (b) ratio of user-intervention time to total task duration for four intervention types and overall. (c) Task-wise and Step-wise Intervention Rates for DroidRetriever. (d-e) Action steps, (f-g) Token Usage, (h-i) Completion Time, (j-k) three objective metrics (Coverage, Accuracy, Redundancy), and (l-m) three subjective metrics (Workload, Certainty, Confidence) for single-app and multi-app tasks. *** indicates a significant difference with *** for p < .001, ** for p < .01, and * at p < .05.}
    \centering
    \includegraphics[width=0.85\linewidth]{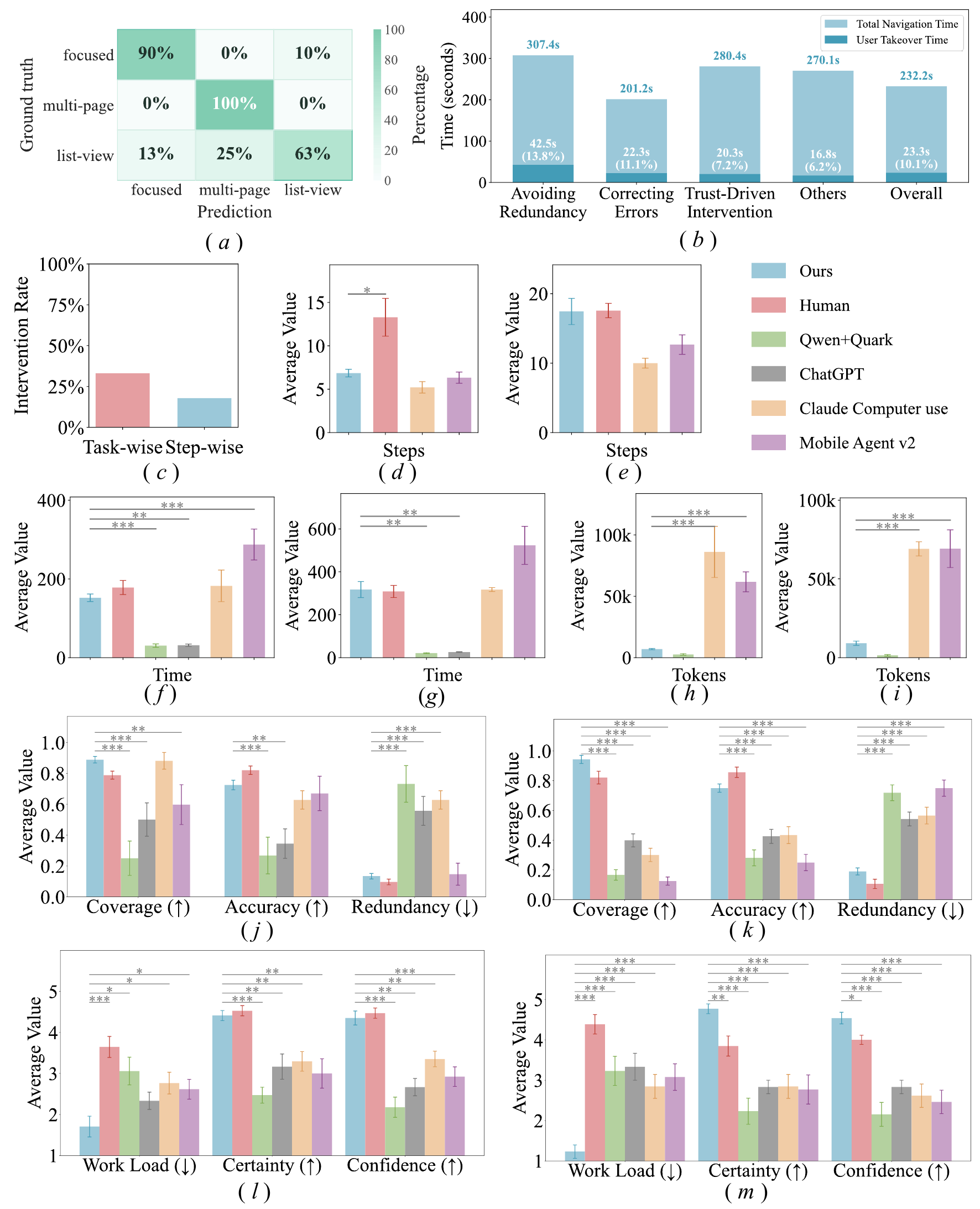}
    \caption{\textcolor{black}{Results of Study 2, including task decomposition and a comparative evaluation of Human, DroidRetriever, LLM-driven search engines (Qwen \& ChatGPT), Claude Computer Use, and Mobile-Agent-v2. (a) Page-level decomposition confusion matrix. (b) Ratio of user-intervention time to total task duration for four intervention types and overall. (c) Task-wise and Step-wise Intervention Rates for DroidRetriever.(d-e) Action steps, (f-g) Token Usage, (h-i) Completion Time, (j-k) three objective metrics (Coverage, Accuracy, Redundancy), and (l-m) three subjective metrics (Workload, Certainty, Confidence) for single-app and multi-app tasks. *** indicates significant difference at \( p < .001 \), ** at \( p < .01 \), and * at \( p < .05 \).}}
    
    \label{fig:study2}
\end{figure*}

Page-level decomposition accuracy across 22 tasks was 82\% (focused: 90\%, multi-page: 100\%, list-view: 63\%). As illustrated in Figure~\ref{fig:study2}(a), performance is strong on focused and multi-page tasks, while list-view tasks are occasionally misclassified as multi-page. This behavior is acceptable in practice: for example, in the “check liked playlists” task, the agent correctly navigates to the liked playlists and saves the results, and may additionally open and save two other playlists, but the final report filters them out. But it is worth noting that, in a few cases, the agent opened each detail page before aggregating candidates at the list level, resulting in fewer collected results.

\begin{table*}[t]
  \centering
  \footnotesize
  \renewcommand{\arraystretch}{1.15}
  \Description{Table 4: Compact mapping between the four open-web failure modes discussed in Section 6.2 and the Study 2 tasks. Each row lists a failure mode on the left and representative tasks on the right. Task IDs use the prefix S for single-app tasks and M for multi-app tasks.}
  \caption{Open-web failure modes and example tasks. (IDs: S-\# for single-app tasks; M-\# for multi-app tasks)}
  \begin{tabular}{p{4cm} p{12cm}}
    \hline
    \textbf{Failure mode} & \textbf{Tasks and IDs} \\
    \hline \hline

    \textbf{Context Barrier} &
    Amap ETA requires user context (S-4),\;
    SHEIN currency settlement units (S-13),\;
    Big Mac ordering decision (M-5),\;
    VIVO X100 Ultra purchase advice (M-6),\;
    Showtimes for \textit{F1: The Movie} (M-7) \\
    \hline

    \textbf{Interaction Barrier} &
    Big Mac ordering decision (M-5),\;
    Showtimes for \textit{F1: The Movie} (M-7) \\
    \hline

    \textbf{Rendering Barrier} &
    King Gnu popular songs on Spotify (M-1),\;
    Showtimes for \textit{F1: The Movie} (M-7) \\
    \hline

    \textbf{Source Inconsistency} &
    Reviews of Marshall Middleton on Rednote (S-6),\;
    King Gnu popular songs on Spotify (M-1),\;
    Orlando Disney travel guides across platforms (M-2) \\
    \hline

  \end{tabular}
  \label{tab:study2-failuremode-map}
\end{table*}
Under fully autonomous, no-intervention settings, navigation accuracy was 77\%. Failures fell into three categories: two tasks with deeply nested that made navigation challenging; two tasks caused by decomposition errors that jumped to detail pages and reduced aggregation; one task reaching a second-level detail page without promotional information, posing a price miscalculation risk.



We observed user-initiated reactive interventions during the study. These reactive interventions arose in four recurrent scenarios: (a) \textbf{Correcting Errors (62.5\%)}: when the system executed erroneous operations, such as mishandling unexpected pop-ups; (b) \textbf{Trust-Driven Intervention (12.5\%)}: when navigation errors occurred that the agent could have recovered from, yet users still intervened due to insufficient trust or familiarity with these capabilities (e.g., when encountering duplicate search results); (c) \textbf{Avoiding Redundancy (10.4\%)}: when the system engaged in redundant actions, like proposing further exploration after participants had already obtained satisfactory results; and (d) \textbf{Others 14.6\%}: including cases such as human-machine verification and network failures. This manifested as a 33.1\% task-level intervention rate (Figure~\ref{fig:study2}(c)). Despite 77\% navigation accuracy, participants still intervened. For tasks requiring intervention, the average number of interventions was 1.837 ($SEM=0.128$), and the average intervention steps needed was 2.571 ($SEM=0.369$). The step-level intervention rate was 17.9\%, suggesting that only a few corrective steps were usually enough to bring DroidRetriever back on track.

Figure~\ref{fig:study2}(b) shows the proportion of user intervention time relative to the total task time across four categories of intervention. Among them, Avoiding Redundancy is the most time-consuming and intervention-intensive type, with the longest overall task completion time. Users need to spend additional time identifying redundant system behaviors, which leads to a noticeable decline in human-AI collaboration efficiency. Correcting Errors ranks second. While simple errors can be quickly detected and resolved, tasks that require multiple operations to restore the interface, such as when the system search with default recommended keywords but not the user required one, demand significantly more user effort and time.


\textcolor{black}{Performance metrics were analyzed using repeated-measures ANOVA with Tukey’s HSD post-hoc tests, while Likert-scale responses were evaluated using Kruskal-Wallis tests followed by Dunn’s post-hoc tests with Holm correction (post-hoc comparisons were conducted only between our method and the other approaches).} As shown in Figure~\ref{fig:study2}(d-e), even in the single-app setting, dispersed information led participants to frequent page switching and the highest step count ($F=4.980, p=.003$). In the multi-app setting, there was no significant difference in step counts between DroidRetriever and human participants ($F=2.220, p=.093$), demonstrating human-like navigation behaviors without introducing additional redundancy. DroidRetriever also showed comparable task completion times to those of participants in both single-app ($F=18.119, p=.869$) and multi-app tasks ($F=10.190, p=1.000$). As shown in Figures~\ref{fig:study2}(j) and (k), DroidRetriever achieved slightly higher coverage than humans (single-app: $F=16.819, p=.070$; multi-app: $F=63.175, p=.449$), similar accuracy (single-app: $F=11.165, p=.209$; multi-app: $F=27.814, p=.605$), and acceptable redundancy levels (single-app: $F=38.465, p=.790$; multi-app: $F=34.029, p=.797$). The minor loss in accuracy mainly occurred in tasks involving multiple arithmetic operations, consistent with the observation in Study 1. As illustrated in Figures~\ref{fig:study2}(j-m), in both the single-app and multi-app scenarios, participants completing the tasks manually reported significantly higher subjective workload than those using DroidRetriever \textcolor{black}{($H=22.916, p<.001; H=34.379, p<.001$)}. In terms of perceived certainty and confidence, no significant differences were observed in the single-app scenario \textcolor{black}{($H=47.817, p=.511; H=51.288, p=.742$)}. In the multi-app scenario, however, participants using DroidRetriever reported significantly higher perceived certainty \textcolor{black}{($H=34.898, p=.003$)} and confidence \textcolor{black}{($H=43.124, p=.011$)} compared to manual completion. These results suggest that DroidRetriever not only reduces users’ cognitive burden, but also helps strengthen users’ trust when tasks involve cross-app information seeking.


In multi-app tasks, the workload during manual completion was higher ($M=4.385, SEM=0.241$) and certainty ($M=3.846, SEM=0.249$) and confidence ($M=4.000, SEM=0.113$) were lower than in single-app conditions (workload: $M=3.647, SEM=0.256$, certainty: $M=4.529, SEM=0.125$,confidence $M=4.471, SEM=0.125$), but the reductions in these two metrics were less pronounced for DroidRetriever (multi-app: workload $M=1.231, SEM=0.166$; certainty $M=4.769, SEM=0.122$; confidence $M=4.538, SEM=0.144$; single-app: workload $M=1.706, SEM=0.254$; certainty $M=4.412, SEM=0.123$; confidence $M=4.353, SEM=0.170$). This may be because in manual operation, participants need to memorize more information points and perform more actions in multi-app tasks, which increases the work load and makes participants experience more difficulties. This also affects perceived confidence level. While under the assistance of DroidRetriever, the whole progress is visualized and easy to trace back. This relieves the participants memory requirement, and leads to a higher confidence level in the result.


The LLM-driven search engine demonstrate advantages in speed and token usage by retrieving information directly through APIs, as shown in Figures~\ref{fig:study2}(f-i). However, compared with DroidRetriever, both LLM-driven search engines exhibited significantly lower coverage and accuracy in both single- and multi-app scenarios (single-app: Qwen: coverage $F=16.819, p<.001$; accuracy $F=11.165, p<.001$; ChatGPT: coverage $F=16.819, p<.001$; accuracy $F=11.165, p<.01$; multi-app: Qwen: coverage $F=63.175, p<.001$; accuracy $F=27.814, p<.001$; ChatGPT: coverage $F=63.175, p<.001$; accuracy $F=27.814, p<.001$). Additionally, when the system retrieves available information, it tends to generate comprehensive and detailed reports, which can lead to a higher redundancy (single-app: Qwen $F=38.465,p<.001$; ChatGPT $F=38.465,p<.001$; multi-app: Qwen $F=34.029,p<.001$; ChatGPT $F=34.029,p<.001$).



The underperformance of LLM-driven agents in these tasks can be attributed to two primary factors. First, their access to platform-specific content and information requiring user authentication is limited. Consequently, they perform particularly poorly on login-gated tasks (as shown in Table~\ref{tab:study2-single-app-task} and Table~\ref{tab:study2-multi-app-task}). Second, even when dealing with open-web content that appears publicly accessible, we observed that general LLM agents still exhibit notable limitations \textcolor{black}{(summarized in Table~\ref{tab:study2-failuremode-map})}: (1) \textbf{Context Barrier}: Certain key information depends on user-specific states such as membership level or historical preferences (e.g., cookies). General-purpose agents lack access to such contextual data, leading to biased or inaccurate outcomes (e.g., wrong prices). (2) \textbf{Interaction Barrier}: Many queries require active web interaction, for example, entering a location to estimate delivery time. Such information is only accessible after specific user input triggers an API call. General LLMs usually cannot execute these interactive steps. (3) \textbf{Rendering Barrier}: Many modern websites adopt client-side rendering (CSR) and rely on JavaScript to dynamically load content (e.g., Spotify artist pages). While humans using browsers automatically execute scripts to render full pages, LLM agents usually capture only the skeletal HTML structure without essential content. (4) \textbf{Source Inconsistency}: When encountering any of the above barriers or anti-scraping mechanisms, general LLMs often silently switch to alternative data sources to assemble an answer without explicitly notifying the user. As a result, users may only discover such inconsistencies upon manually verifying the referenced links.

Figure~\ref{fig:study2}(d-i, j and k) show results across six objective metrics. Although Claude computer use required the fewest average steps, it still took long average time especially in multi-app tasks. This is primarily due to two factors. First, Claude performs multiple rounds of internal deliberation for each operation, increasing the number of calls to the LLM. Second, it saves the last three screenshots and the entire textual history as input for the model, significantly increasing both computational load and time costs. On average, Claude consumed approximately 210,128 tokens per task, significantly higher than other systems. Notably, our system reduced token usage to 19,459 tokens by using small models to convert UI information into natural-language descriptions while also accelerating processing compared with other mobile agents.

However, the high cost of Claude did not bring better performance, particularly in accuracy and redundancy, even with extensive user interventions. We further verified that Claude’s internal reasoning text corresponded exactly to its output click positions, ruling out potential resolution discrepancies caused by screen mirroring. A key issue was Claude's memory mechanism, which stored all interactions without distinguishing report-relevant information from navigation details, which contaminates the final report with incorrect information and loses part of the history of visited pages. For tasks involving multiple apps and pages (Table~\ref{tab:study2-multi-app-task} tasks 2, 5, 6), Claude often terminated after collecting information from only a single app or page, leading to incomplete results. Because Claude presents outputs as a dialogue stream containing its reasoning history, participants found it difficult to follow its progress. Combined with navigation failures, lack of proactive scrolling within target interfaces (Table~\ref{tab:study2-multi-app-task}: tasks 2, 6, 8), and reporting errors, even with user intervention, Claude showed significantly lower coverage ($F=63.175,p<.001$) and accuracy ($F=27.814,p<.001$) in multi-app tasks compared with DroidRetriever (Figure~\ref{fig:study2}(j-k)).



Similarly, Mobile-Agent-v2 is optimized for task automation, leading to problems in multi-source information gathering (dialog drift and premature termination, Table~\ref{tab:study2-multi-app-task} tasks 2, 6, 8) and information retention (forcing users to manually verify or relying on limited memory that yields fragmented information, Table~\ref{tab:study2-multi-app-task} tasks 2, 5). Even though it includes a memory mechanism tailored for multi-app tasks, our inspection shows that this memory is an LLM’s secondary summarization of screens, which omits key details and cannot reliably link reports back to specific UI content. Consequently, it achieved significantly lower coverage (single-app: $F=16.819, p<.001$; multi-app: $F=63.175, p<.001$) and lower confidence \textcolor{black}{(single-app: $H=51.288, p<.001$; multi-app: $H=43.124, p<.001$) as shown in Figure~\ref{fig:study2}(j-m)}. Mobile-Agent-v2 lack effective user intervention, leading to conflicts over device control between users and agents.

\section{Study 3: Case Study}  
Encouraged by the findings from the previous two studies, we conducted a case study to take a closer look at the usefulness and effectiveness of DroidRetriever. We focused on four questions: (1) the overall experience of our system compared with other methods, (2) the quality and accuracy of the reports, (3) how the system supports transparency and trust, and (4) the experience of taking over and sense of control.

\subsection{Method}
In this study, we recruited 13 new participants who had not taken part in Study 2, including 4 professionals with experience in mobile agents and 9 regular users (ages 22-33, 2 female). Participants first selected one single-app task and one cross-app task from the 22 tasks used in Study 2. Both tasks had to be topics they intended to explore but did not already have the answer. For each task, participants then created four variants of similar difficulty, for example by switching the app used, modifying the search terms or filtering criteria, or varying the format and emphasis of the output. These variants also had to be tasks they were interested in without obvious answers.

The study was conducted under five conditions: manual search, LLM-driven search agents, Claude Computer Use, Mobile-Agent-v2, and our system. All system implementations were identical to those in Study 2. Each participant completed two sets of tasks, and within each set, every task was performed under all five conditions. The order of conditions was randomized. Participants received a 5-10 minute tutorial and hands-on practice with all the systems before the study. Participants were free to check and intervene at any time using their preferred methods to ensure the quality of the report. We encouraged participants to think aloud during takeover, which helped us better understand the boundaries of control and the mechanisms of trust. The study concluded with a semi-structured interview that focused on four aspects: participants’ comparative experiences across methods, their perceptions of the quality and accuracy of system reports, how transparency and trust were established, and their sense of control during takeovers. The full interview questions are included in Appendix~\ref{appendix:study3-questionnaire}. Participants were compensated with \$10 (USD).

\subsection{Results}
We organize the results into four qualitative findings from post-study interviews and a summary of observed takeover behaviors.

\subsubsection{Overall experience compared with other methods}

Overall, participants found DroidRetriever navigated to the correct pages, aligning with our Study 1 findings (Sec.~\ref{sec:study-1}). As P2 put it, "\textit{Only DroidRetriever actually completes multi-app tasks, while the other systems’ reports are hard for me to actually use}". LLM-based web search often "\textit{missed key points}" in consolidation (P13) and could not access information on closed, login-gated platforms (P10). Participants also noted limitations of the two mobile agents: "\textit{Claude and Mobile-Agent-v2 produce reports that are incomplete for multi-app tasks without the original screens. Sometimes they just wander within a single app, so I can’t trust the results}" (P3). Others reported that "\textit{Mobile-Agent-v2 often only opens pages without answering my question, so I have to inspect everything myself}" (P6), and that its multilingual capability was weak (P2).

DroidRetriever was considered more intuitive to interact with. Two participants emphasized the necessity of voice input: "\textit{if voice is not supported, the cognitive load increases}" (P4). Most participants preferred taking over by directly operating the phone, describing it as more natural and efficient: "\textit{the interaction logic matches my habits, and next-step visualization is clear}" (P5); "\textit{compared to issuing natural language commands, direct manipulation avoids having to formulate instructions and negotiate with the agent}" (P9); "\textit{it’s like rescuing a vacuum stuck in a corner, and once I nudge it closer to the information entry, most navigation errors can be corrected}" (P3). By contrast, Claude’s presentation of the information gathering process was viewed as less user-friendly: "\textit{Claude’s interaction is not intuitive, and the flow is too complex}" (P7); "\textit{the chat window keeps bubbling up messages, making it hard to follow its progress}" (P3).

Participants noted that manual work becomes taxing for cross-app comparison, browsing and summarizing long or multi-page content, and cases needing domain knowledge (P2/5/6). They also affirmed the system’s value in easing these challenges: "\textit{we need a fast and reliable assistant for tasks that span several apps and end with a summary}" (P5).

\subsubsection{Report quality}

Participants agreed the information retrieved by our method was up-to-date and generally accurate. However, shopping scenarios still showed issues: "\textit{some hidden deals may be missed}" (P7); "\textit{the results match my preferences but do not fully account for member discounts}" (P9); "\textit{price calculations are sometimes inaccurate}" (P13). One participant also noted: "\textit{in Claude and LLM search engines I cannot directly compare and trace original sources, which is cumbersome and reduces my trust}" (P8).

Participants consistently noted that information sources directly affect accuracy and trust regardless of the system. Some argued that adding more sources helps mitigate single-source bias (e.g., soft ads) (P7/8/9); others cautioned that "\textit{agents may blur differences among authors’ viewpoints when aggregating content}" (P7). Participants also emphasized that retaining and clearly labeling citations in reports is crucial for traceability and verification (P4/11/7).

Participants reported that DroidRetriever provided more complete information: "\textit{It can retrieve sufficient content under my specified search depth. Even if the report omits details, I can still view the original screens}" (P3). By contrast, "\textit{LLM outputs were obviously incomplete and sometimes not from my requested sources}" (P13), and "\textit{Claude’s coverage was insufficient, with no scrolling operations at result pages}" (P1). When encountering omissions, most participants (12 out of 13) preferred to search manually or provide manually retrieved sources for agent to integrate (P4); only one participant (P11) continued to prompt the agent to refine the answer. These requirements indicate that report synthesis module is useful as a standalone capability.

Participants noted that redundancy hindered reading efficiency, especially in LLM-driven search agents (P4/13), though many had adapted to quickly skipping repeated content (P2/4/7/8/12/13). Reported causes of redundancy included "suboptimal result page selection" (P7), "incomplete understanding of user intent" (P4), and insufficient result filtering mechanisms (P8/12). Nearly half preferred to view the report first with original interfaces as backup (6 out of 13), while others preferred direct access to original interfaces for reference. Overall, retaining original interfaces was seen as necessary: "\textit{I prefer to see the report content with links to original screens for further verification, and DroidRetriever matches my expectations}" (P5); "\textit{I prefer original interfaces because LLM summaries are not always reliable}" (P6).

\subsubsection{Transparency and trust}

All emphasized that the clarity of information sources was the primary factor shaping their trust. They consistently expected the system to explicitly indicate the origin of key evidence and conclusions, ideally pointing concisely to specific platforms, merchants, or authors. One participant noted: "\textit{ultimately my decisions are based on the references I can directly inspect, and the system’s summary is only auxiliary}" (P7). Another stressed: "\textit{the credibility of the information source itself is the most critical transparency indicator}" (P9), while a third observed: "\textit{unreliable or untrustworthy sources directly affect my judgment}" (P8). Overall, the accuracy and transparency of the sources used in reports proved decisive for fostering users’ trust.


Most participants (11/13) found DroidRetriever’s progress dashboard clear and understandable, as it provided a transparent view of the agent’s traces and helped them judge whether the system remained on track. Professionals tended to seek system-level recovery strategies, such as rollback (P2), rerun, or clearing background tasks (P7), focusing on whether the agent could "\textit{continue running}" (P2). In contrast, non-professionals relied more on the process transparency conveyed by the interface to build understanding and trust: when familiar with the target application, they could quickly identify and correct errors (P1/3/5), but in unfamiliar apps they "\textit{need to consult the operation trace to locate problems}" (P5). As one participant explained: "\textit{it is difficult for me to find potential mistakes made by Claude without constant monitoring; DroidRetriever’s dashboard is far more user-friendly so I know it doing exactly what I asked}" (P12). Process transparency across the interfaces became a crucial factor in sustaining users’ trust in the system’s operation.

\subsubsection{Intervention and sense of control}

Participants widely regarded task progress visualization and the ability to take over at any moment as the most valuable features (12 out of 13). These functions helped them quickly understand the agent’s status and strengthened their sense of control. Some participants said the progress dashboard showed the agent’s reasoning across multi-step or cross-platform tasks, making it easier to decide when to intervene (P5/9/10). Others emphasized its value for information-intensive tasks, noting they could "\textit{let the agent run on its own, and only occasionally check whether it remains on the right track}" before stepping in if needed (P3). Several also stressed that takeover was not only for fixing mistakes but for knowing they could "\textit{interrupt and regain control at any time}", which eased concerns about relying too much on the agent (P4/11). P10 added that the design "\textit{lets me decide whether to hand tasks back to the agent when sensitive or private information is involved}", which enhanced their sense of control and security. Overall, progress visualization and takeover were consistently recognized as central to user control and confidence.

Most participants found the current task checklist and tick-off display clear enough to track the agent and strengthen their sense of control (P1/2/3). For risk alerts, they preferred immediate cues such as vibration, sound, or pop-ups (P6/7/9/10). They also wanted the option to pre-configure notification modes, or in some cases skip alerts and allow automatic execution (P3).

On privacy boundaries, they agreed that account passwords, verification codes, and payment operations must remain under direct user control and should never be automated by the agent (P1/2/6/10). Many also viewed "\textit{money transfers and message sending as high-risk tasks, preferring to perform them manually or confirm step by step since errors would be costly to fix} (P3). For data such as location, browsing history, or in-app contacts, some felt that one-time authorization was sufficient and repeated prompts unnecessary (P1/3). Others noted that their willingness to authorize depended on system reliability:  "\textit{if the system proved trustworthy, sharing routine identifiers such as phone numbers or names was acceptable, but payment credentials remained a strict boundary}" (P5/9). Overall, participants wanted strict boundaries but also flexible authorizations to balance privacy and convenience.

\section{Discussion}
We discuss the design of proactive takeover prompts, progress visualization design, potential application and limitations in this section.

\subsection{Adaptive Takeover Notifications}

Since current mobile agents still fall short of fully and reliably completing long-horizon, multi-app tasks, appropriate notifications are important  when requiring human assistance, e.g., state confirmation or small corrections at key steps. In particular, when the agent proactively pauses and ask for user confirmation, it should leverage the phone’s native OS and ecosystem to deliver diverse takeover notifications that fit different contexts and user groups.
For instance, in quiet environments or for users who find ringtones awkward in public, vibration or in-ear cues are preferable to speaker output; during driving or other enclosed, hands-busy scenarios, voice interaction is critical; and for users with hearing impairment or in noisy environments, smartwatch vibrations are more accessible. If the system can automatically infer the user’s context and adapt its notification strategy, users can be notified more promptly and overall experience will be improved.

\subsection{Design for System Transparency}
We enhance system transparency in two ways to strengthen user control: (1) a task progress dashboard; (2) original screenshots with traceable citations in the report.

Many agents only provide lengthy and sequential chat logs or video replays, or even no history at all. We design and implement a navigation progress dashboard designed within our task decomposition paradigm. It provides a clear visual presentation of the exploration map, describing the process from target apps to search terms, and sequences of page transitions. Inspired by debugging workflows, we place a blue highlight around the agent’s current screen to indicate its position on the exploration map. We use green boxes to mark pages stored in the search results database and treat these pages as milestone nodes. This makes it easy for users to revisit key progress and intermediate outputs. In study 3, participants endorsed this design. They found it intuitive, easy to understand and it increased their trust when letting the agent run without continuous supervision.

To further bolster user confidence in the generated reports, we retain original screenshots and highlight key information sources within them. Whenever critical details appear in a report, there will be a link to these screenshots, allowing users to verify the information. This referencing mechanism ensures that even if large language models produce inaccuracies or misunderstand interface elements, users can still validate the accuracy of the information and get accurate information, which again enhances their trust in the reports. This transparent and verifiable design principle not only improves user experience but also fosters positive interaction between the system and users.

\subsection{System Value and Motivation}
As an independent capability, report synthesis offers clear user value. Not all users expect agents to automatically navigate to information pages; many prefer to control their own browsing pace and simply need concise summaries or screen sketches
to aid comprehension and quick recall. 
For example, when 
browsing apps and web pages (e.g., news, product pages, privacy policies), users often need lightweight summarization and just-in-time Q\&A. Traditional screenshots only capture whole pictures and lack semantic analysis and organization, while report synthesis can provide task-oriented distillation and structured results based on user-selected content (e.g., a long screenshot or multiple screenshots). This reshapes user interactions with the interface and supports further downstream decision-making and archiving. Moreover, if such reports support long-term storage and retrieval, they can serve as durable personal reference storage with the potential to evolve into light-weight knowledge notes. 
The structured format also significantly enhances sharing and collaboration efficiency. It empowers teams or family members to effortlessly synchronize and stay aligned on crucial information.


DroidRetriever targets everyday, just-in-time information-seeking scenarios. For example, while working with the phone set aside, users may want the agent to answer questions based on real-time data from installed apps, such as “What should I have for dinner tonight?” or “How should I plan next week’s trip?” However, general-purpose LLM search engines often struggle in these specific contexts. They lack access to implicit user context (such as location-based pricing), cannot perform complex interactions required to fetch data, fail to render JavaScript-heavy content on modern platforms, and may silently substitute blocked sources with inconsistent alternatives. Therefore, DroidRetriever is designed for these high-frequency, day-to-day use cases and prioritize timely, actionable responses. It utilizes installed apps, which are familiar and frequently used for users. This reduces learning costs and helps build user trust in the system.

\subsection{Limitation and Future Work}

Despite demonstrated capabilities in information retrieval and navigation, DroidRetriever faces limitations particularly with dynamic interfaces. Relying on screenshots and screen recognition, the system struggles with dynamic contents, such as video streams, which hinders its ability to summarize and integrate information from multiple video sources. Additionally, delays between capturing screenshots and executing operations may impede timely responses to sudden screen changes, such as intrusive ads. 
Future work should prioritize enhancing support for dynamic interfaces and improving the system's responsiveness to screen changes.

Our system, built on a generalized large language model, is not specifically optimized for mobile application execution. Consequently, while we aim to provide clear screen understanding and design mechanisms to minimize navigation errors, the system's performance still relies on the planning capabilities of the utilized LLM. Furthermore, the phased understanding and decision-making process introduces notable latency. Future research could explore employing multi-modal large models, such as CogAgent~\cite{Cogagent2024}, specifically trained for mobile interfaces to facilitate end-to-end navigation decisions and improve speed.

\section{Conclusion}


This paper introduces DroidRetriever, a transparent, steerable mobile information seeking system based on multi-LLM collaboration. DroidRetriever takes voice or text queries, navigates to relevant app screens, captures screenshots, extracts and integrates information, and delivers a concise report with citation-linked screenshots. The system comprises three modules: task decomposition, UI copilot, and report synthesis. It automatically selects candidate applications, breaks down tasks into sub-tasks, and executes step-by-step navigation. During navigation, it provides a real-time progress dashboard that surfaces sub-task status and live exploration traces that users can check at any time. Ultimately, the system synthesizes a report with precise citations. 

Our user study shows the efficiency and accuracy of our approach in report synthesis, with participants highly rating the report quality. Another study involving 22 real-world tasks (14 single app tasks and 8 multi-app tasks) across four categories (summarization, comparison, processing, and localization) demonstrated consistent improvements over existing tools in report quality, result experience, and perceived trust. Through in-depth semi-structured interviews, we further characterized users’ trust in the design and their considerations around takeover strategies.

\begin{acks}
We sincerely thank Professor Xiaohong Guan for his generous support and guidance throughout this work. This work is supported by Xiaomi Open-Competition Research Program.
\end{acks}

\bibliographystyle{ACM-Reference-Format}
\bibliography{chi26ref}

\newpage

\appendix

\section{Appendix: Illustration of Scrolling Screenshot}\label{appendix:scrolling-screenshot}

\begin{figure}[H]
    \centering
    \includegraphics[width=0.9\linewidth]{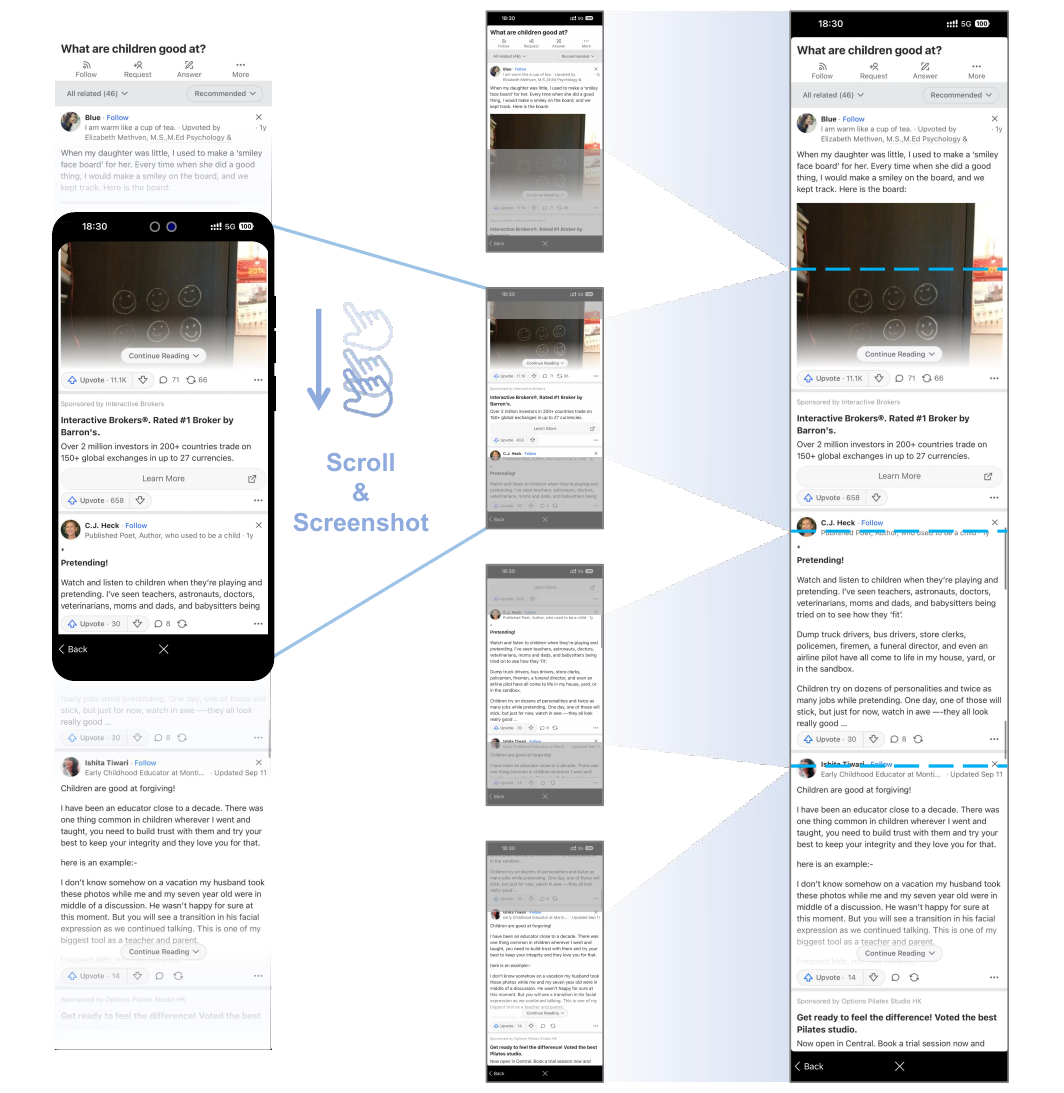}
    \Description[Illustration of Scrolling Screenshot]{Illustration of Scrolling Screenshot}
    \caption{Illustration of scrolling screenshot.}
    \label{fig:scrollingscrshot}
\end{figure}

As shown in Figure~\ref{fig:scrollingscrshot}, the "scrolling screenshot" operation involves performing multiple downward scrolls and capturing several screen regions. The system then uses template matching to stitch these segments into a single long-page image.

\section{Appendix: LLM Prompts in DroidRetriever}

\subsection{LLM1: Task Decomposition}

\textbf{System Instruction:} \\
Please answer the following questions:
\begin{itemize}
    \item Extract the app names explicitly mentioned in the task.
    \item List apps that are installed and relevant to the task (up to 3).
    \item List apps that are not installed but relevant to the task (up to 3).
    \item If a query is needed, provide up to 3 search terms.
    \item Select the query mode: multi-page, focused, or list-view.
\end{itemize}

The task requirement is \textbf{\{task\_name\}}, and the following apps are installed: \textbf{\{app\_list\}}.

\textbf{Sample output format:}
\begin{verbatim}
{
    "mentioned_apps": [Expedia, Booking],
    "installed_related_apps": [Expedia, Booking],
    "uninstalled_related_apps": [none],
    "search terms": ['Universal Studios Japan'],
    "search_mode": ['Multi-page']
}
\end{verbatim}

\subsection{LLM2: UI Navigator}

\textbf{System Instruction:} \\
You need to act as a smartphone assistant:
I need to complete a task on a mobile app but am unsure how to proceed. Please tell me which element to tap or what content to enter based on the task, the controls I've tapped, and what I've entered on the keyboard.

If I provide help document information, please refer to it first, but also take into account the actual interface, focusing on the real buttons. The interface I provide may not be the initial one, as some actions might have already been completed. Based on this, please determine the next step and provide a standardized operation command.

Q: \textbf{\{sub\_task\_query\}}, previous actions: \textbf{\{previous\_actions\}} \\
Current screenshot contains the following contents: \\
\textbf{\{current\_screen\}} \\
You can refer to this help document: \textbf{\{help\_document\}}

\textbf{Sample output format:}
\begin{verbatim}
{
    "action": "tap",
    "tap_point": [535, 1490],
    "element_location": {"left": 475, "right": 595, 
                         "top": 1430, "bottom": 1550}
}
\end{verbatim}

\subsection{LLM3: UI Evaluator}

\textbf{System Instruction:} \\
You are a dual-role assistant. Based on the interface and actions, please judge both:
\begin{enumerate}
    \item[(1)] Whether the current app task is complete.
    \item[(2)] Whether the current screen belongs to a high-risk or privacy-sensitive interface requiring manual user takeover.
\end{enumerate}

\textbf{Task Completion Criteria}
\begin{itemize}
    \item For "list" mode: The task is considered complete once navigation to the search results page is achieved.
    \item For "focused" mode and "multi-page" mode: The task is only considered complete when navigation to the details page of the search results is achieved.
    \item Hotel Search Task: The screen must show specific room prices for one hotel. Listing multiple hotels doesn't count.
    \item Shopping Task: Completion requires reaching the product page with options like "Customer Service," "Favorites," or "Shopping Cart."
    \item For tasks involving specific content, completion is only achieved when the action of clicking on the article title is performed.
\end{itemize}

\textbf{Current task:} \{sub\_task\_query\} 
\textbf{Following actions have been performed:} \{previous\_actions\}
\textbf{Current screen:} \{current\_screen\}

\textbf{Risk / Privacy-Sensitive Screen Criteria}
\begin{enumerate}
    \item Login/Registration/Identity verification (account, password, verification code, ID, face/fingerprint recognition).
    \item Payment/Transfer/Recharge/Withdrawal (payment password, bank card, code, biometric, order submission, purchase confirmation).
    \item Personal information editing (name, ID number, address, birthday, ID photo, etc.).
    \item System/App privacy settings (permissions such as location, contacts, camera, microphone; factory reset; delete all data; change security settings).
    \item Account deletion/data removal (cancel account, delete data, clear chat history, irreversible actions).
    \item Agreement/Authorization interface (consent to privacy policy, user agreement, third-party login authorization).
    \item Sensitive professional apps (medical, legal, government) involving uploading/filling sensitive information.
\end{enumerate}
\textbf{Not high-risk:} Viewing ordinary settings (Wi-Fi, Bluetooth, brightness, sound), normal browsing/searching, chat/mail view/edit, local photo/file browsing. If the task itself requires private info lookup, it does not count as high-risk.

\textbf{Output Format}
\begin{verbatim}
Completion<start>[True/False]<end>
Reason<start>[Task completion reasoning]\end
Risk<start>[True/False]<end>
Reason<start>[If True, specify which high-risk criteria; 
              if False, explain why not]\end
\end{verbatim}

\subsection{LLM4: Report Synthesis}

\textbf{System Instruction:}
\textbf{Character Setting and Task:}
You are now a well-trained interface information extraction and integration robot, capable of strictly following my requirements to answer questions without accessing additional information online.
You need to extract, summarize, or integrate content based on the text information from \textbf{all interfaces} I provide, and select and return \textbf{different report formats} according to \textbf{different task types}.

The specific requirements are as follows:

\textbf{Citation Requirements}
\begin{itemize}
    \item Each key point in the answer must be annotated with the source of the search results. The citation format is: [x(interface original content)].
    \item Here, x is the \textbf{interface id} (not the line number), and "interface original content" refers to the \textbf{specific element's original text} on the interface referenced for the key point. If there are multiple citations, use multiple brackets, e.g., [[1(xxx)][2(yyy)]].
    \item Provide citation sources for as many key points as possible.
\end{itemize}

\textbf{Task Types}
\begin{enumerate}
    \item \textbf{Article Summary}: You need to combine one or more interfaces to summarize and provide a relatively reasonable summary of the article's key points. For example: However, some users expressed dissatisfaction with this song[3(not good)].
    \item \textbf{Comparison Task}: You need to combine one or more interfaces to provide a comparison from multiple perspectives in the form of a \textbf{markdown table}, based solely on the given information. For example, for the task "Compare the performance of iPhone 14 and 14 Pro," you need to compare camera parameters, screen size, weight, etc. Note that all comparison information must be explicitly provided on the interface, e.g., price 120 yuan[1(120)], weight 450g[2(450g small capacity)].
\end{enumerate}

\textbf{Task: }The task I need to complete now is: \textbf{\{task\_name\}}. Please refer to the following multiple interfaces and answer in the required format.
\textbf{Citations are mandatory}:
\textbf{\{scr\_info\}}

\textbf{The output must be in markdown format. Citations are mandatory.}

\section{Appendix: Questions for the Interviews}
\subsection{Details of Study 1} 
\subsubsection{Questionnaire for Ratings} \label{appendix:study1-questionnaire}
\begin{enumerate}
    \item How difficult was it for you to complete the task? 
    \begin{itemize}
        \item $\square$ Simple (The task is straightforward and can be done with minimal effort or assistance.)
        \item $\square$ Moderate (The task has a moderate degree of complexity and requires several steps or a bit of thought to finish.)
        \item $\square$ Difficult (The task is complex, involving multiple steps, in-depth analysis, or specialized knowledge to complete.)
    \end{itemize}
    
    \item Do you think the information points provided in the report are accurate? \\
    (\textcolor{black}{1} = Not accurate at all, 5 = Completely accurate) 

    \item To what extent does the report cover the information needed to solve the task? \\
    (\textcolor{black}{1} = Covers very little, 5 = Fully covers all necessary information) 

    \item How easy is it to read and understand the report? \\
    (\textcolor{black}{1} = Very hard to read, 5 = Very easy to read) 
\end{enumerate}

\subsubsection{Scoring Points} \label{appendix:study1-scoring-points}

The detailed scoring criteria are presented in Table \ref{tab:study1-task-scoring-points}.


\begin{table*}[t]
    \centering
    \footnotesize
    \renewcommand{\arraystretch}{1.2}
    \Description{Table 5: Table titled “Study 1 Tasks: Capabilities, Scoring Points Count, and Criteria.” It lists 13 tasks, each with a capability category, a scoring points count (SPC), and the criteria for earning those points. Tasks 1–3 are Summarization: Task 1 (SPC 16) asks to list Alipay services with password-free or auto-pay enabled by naming 16 apps; Task 2 (SPC 10) summarizes how to use the GTD method on Zhihu by listing the five GTD steps and correctly explaining each; Task 3 (SPC 8) checks the 12306 ticket refund policy by identifying all eight rules. Tasks 4–5 are Comparison: Task 4 (SPC 12) compares OPPO Find X7 Ultra versus VIVO X100 Ultra specifications, requiring correct model-matched specs-battery capacity, charging power, front and rear cameras, body dimensions-and a comparative analysis for each; Task 5 (SPC 6) compares Xiaomi 14 256GB prices and deals on Taobao versus JD, requiring correct model-matched price, trade-in subsidy, and membership discount. Tasks 6–10 are Processing: Task 6 (SPC 20) lists available afternoon trains from A to B on September 12 with four solutions including departure or train number, arrival time, and both stations; Task 7 (SPC 12) shows taxi trips from May to August costing 15–20 yuan by listing three valid records with date, time, origin, and destination; Task 8 (SPC 3) lists delivered packages by express station, classified by two collection stations, with pickup codes in ascending order and all codes correct; Task 9 (SPC 3) identifies the top three most frequent movies across ranking charts, with The Shawshank Redemption as rank 1 and any two from Interstellar, The Legend of 1900, Forrest Gump, Schindler’s List, or The Godfather (1972); Task 10 (SPC 2) calculates phone credit from October to August with the monthly average by providing the correct total or all records and computing the correct average. Tasks 11–13 are Localization: Task 11 (SPC 10) explains functions of nutrients in a baby formula by correctly listing five nutrient categories and their functions; Task 12 (SPC 4) translates a technical post and three replies into English, summarizing the author’s intended message; Task 13 (SPC 2) describes how to disable private messages on Quora by naming the entry “Who can send you message?” and identifying the “No one” option to turn it off.}
    \caption{Study 1 Tasks: Capabilities, Scoring Points Count, and Criteria.}
    {
    \begin{tabular}{c p{5.5cm} c c p{7cm}}
    \toprule
    \# & Task & Capabilities & SPC & Scoring Points Criteria \\
    \hline \hline
    1 & List Alipay services with password-free or auto-pay enabled. & Summarization & 16 & List 16 application names. \\ \hline
    2 & Summarize how to use the GTD work method on Zhihu. & Summarization & 10 & List the 5 GTD steps, and correctly explain each step. \\ \hline
    3 & Check the ticket refund policy on 12306. & Summarization & 8 & Identify all 8 refund policy rules. \\ \hline
    4 & Compare OPPO Find X7 Ultra vs. VIVO X100 Ultra specs. & Comparison & 12 & Provide correct key specifications (must match correct phone models): battery capacity, charging power, front and rear camera specs, body dimensions; give comparative analysis for each parameter. \\ \hline
    5 & Compare Xiaomi 14 256GB prices and deals on Taobao vs JD. & Comparison & 6 & Provide correct key information (must match correct phone model): price, trade-in subsidy, platform membership discount. \\ \hline
    6 & List available afternoon trains from A to B on Sept 12. & Processing & 20 & List four solutions with departure time or train number, arrival time, departure station, and arrival station. \\ \hline
    7 & Show me taxi trips from May to August costing 15-20 yuan. & Processing & 12 & List three valid records with date, time, start point, and destination. \\ \hline
    8 & List delivered packages by express station. & Processing & 3 & (1) Classify by two collection stations; (2) List pickup codes in ascending order; (3) Ensure all pickup codes are listed correctly. \\ \hline
    9 & Identify top 3 most frequent movies across ranking charts. & Processing & 3 & (1) Identify \textit{The Shawshank Redemption} as rank 1; (2) From \{\textit{Interstellar}, \textit{The Legend of 1900}, \textit{Forrest Gump}, \textit{Schindler's List}, \textit{The Godfather (1972)}\}, correctly identify any 2. \\ \hline
    10 & Calculate phone credit from Oct to Aug with monthly average. & Processing & 2 & Provide the correct total (or list all phone credit records) and calculate the correct average. \\ \hline
    11 & Explain the functions of the nutrients in this baby formula. & Localization & 10 & Correctly list 5 categories of nutrients and their corresponding functions. \\ \hline
    12 & Translate the technical post and discussions below into English. & Localization & 4 & Translate the post itself (summarize the author’s intended message) and 3 replies. \\ \hline
    13 & How to disable private messages on Quora? & Localization & 2 & Provide the entry name (\textit{Who can send you message?}) and identify the option to turn it off (\textit{No one}). \\ \hline
    \bottomrule
    \end{tabular}
    }
\label{tab:study1-task-scoring-points}
\end{table*}

\subsection{Details of Study 2} 
\subsubsection{Questions for Interviews} \label{appendix:study2-questionnaire}



\begin{enumerate}
    \item After using this system, I have a clearer understanding of the information needed to complete the task. \\
    \textbf{Scale:} \\
    1 - Strongly disagree; 2 - Partially disagree; 3 - Neutral; 4 - Partially agree; 5 - Strongly agree

    \item How much mental effort was required to complete the task? \\
    \textbf{Scale:} \\
    1 - Very little; 2 - Slightly; 3 - Moderate; 4 - Quite a bit; 5 - A great deal

    \item After completing the task, how confident are you in the decisions you made? \\
    \textbf{Scale:} \\
    1 - Not confident at all; 2 - Slightly confident; 3 - Neutral; 4 - Fairly confident; 5 - Very confident
\end{enumerate}

\subsubsection{Scoring Points} \label{appendix:study2-scoring-points}

The scoring points for single-app and multi-app tasks are detailed in Table \ref{tab:study2-single-app-task-scoring-points} and Table \ref{tab:study2-multi-app-task-scoring-points}.



\begin{table*}
    \centering
    \footnotesize
    \renewcommand{\arraystretch}{1.2}
    \Description{Table 6: Table titled “Study 2 Tasks: Single-App Capabilities, Scoring Points Count, and Criteria.” It lists 14 single-app tasks, each mapped to a capability category, an SPC value, and the scoring criteria. Tasks 1–6 are Summarization: Task 1 (SPC 9) checks which permissions are authorized to Meituan by providing the status of nine permissions; Task 2 (SPC 6) lists Bilibili community guidelines by naming Serious Creation, Friendly Communication, and Embrace Innovation with reasonable explanations; Task 3 (SPC 4) lists 12306 member ticket redemption rights including points-for-ticket redemption, rescheduling rules, and fee standards with two detailed points; Task 4 (SPC 1) checks driving time to Shanghai on Amap, expecting 15 hours within ±1 hour; Task 5 (SPC 3) summarizes three main points from Zhihu reviews of “Black Myth”; Task 6 (SPC 3) summarizes three main points from Rednote reviews of the Marshall Middleton speaker. Tasks 7–10 are Processing: Task 7 (SPC 5) reports how many Eason songs are saved on QQ Music by correctly listing five songs; Task 8 (SPC 3) finds meetings scheduled by Irene, identifying three meetings with details; Task 9 (SPC 1) gives the current tuition and fees on Mobile Campus filtered by given conditions; Task 10 (SPC 1) provides the correct total ride expenses on Amap for last month. Tasks 11–14 are Localization: Task 11 (SPC 10) checks features of the most recently added monitor in a JD cart, providing parameters and explanations including panel type, refresh rate with response time, color gamut, number of colors, and contrast ratio; Task 12 (SPC 5) gives RAM specifications for an item in a Taobao cart, including DDR5, frequency, XMP 3.0 support, timing, and memory model; Task 13 (SPC 4) lists default currency settlement units on SHEIN-USD, EUR, GBP, and CAD; Task 14 (SPC 2) translates a Red Velvet notification from Weverse into Chinese, covering both the time information and event content.}
    \caption{Study 2 Tasks: Single-App Capabilities, Scoring Points Count, and Criteria.}
    {
    \begin{tabular}{c p{5.5cm} c c p{7cm}}
    \toprule
    \# & Task & Capabilities & SPC & Scoring Points Criteria \\
    \hline \hline
    1 & Check which permissions have been authorized to Meituan. & Summarization & 9 & Provide the status of 9 permissions. \\ \hline
    2 & List community guidelines on Bilibili. & Summarization & 6 & List 3 community rule names (\textit{Serious Creation}, \textit{Friendly Communication}, \textit{Embrace Innovation}) and give reasonable explanations for each. \\ \hline
    3 & List ticket redemption rights for 12306 members. & Summarization & 4 & Provide correct information on: (1) points-for-ticket redemption, (2) rescheduling rules, (3) rescheduling fee standards (including 2 detailed points). \\ \hline
    4 & Check the driving time to Shanghai using Amap. & Summarization & 1 & Give 15h (allowable error within ±1h). \\ \hline
    5 & Find and summarize reviews of ``Black Myth'' on Zhihu. & Summarization & 3 & Summarize 3 main points reasonably based on the information found. \\ \hline
    6 & Find and summarize reviews of Marshall Middleton on Rednote. & Summarization & 3 & Summarize 3 main points reasonably based on the information found. \\ \hline
    7 & Find out how many Eason songs I've saved on QQ Music. & Processing & 5 & Correctly list 5 songs. \\ \hline
    8 & Find meetings scheduled by Irene. & Processing & 3 & Correctly identify 3 meetings with detailed information. \\ \hline
    9 & What are my current tuition and fees on Mobile Campus? & Processing & 1 & Provide the correct fee amount, filtered according to given conditions. \\ \hline
    10 & Show me my total ride expenses on Amap for last month. & Processing & 1 & Provide the correct total calculation. \\ \hline
    11 & Check the features of the latest added monitor in my JD cart. & Localization & 10 & Correctly provide monitor parameters and explanations: panel type, refresh rate + response time, color gamut, number of colors, contrast ratio. \\ \hline
    12 & Tell me the specifications of the RAM in my Taobao cart? & Localization & 5 & Correctly provide RAM specifications and explanations: DDR5, frequency, XMP 3.0 support, timing, memory model. \\ \hline
    13 & List all default currency settlement units on SHEIN. & Localization & 4 & Provide 4 units: USD, EUR, GBP, CAD. \\ \hline
    14 & Translate Red Velvet notification from Weverse to Chinese. & Localization & 2 & Correctly translate both the time information and event content. \\
    \bottomrule
    
    \hline
    \bottomrule
    \end{tabular}
    }
\label{tab:study2-single-app-task-scoring-points}
\end{table*}

\begin{table*}
    \centering
    \footnotesize
    \renewcommand{\arraystretch}{1.2}
    \Description{Table 7: Table titled “Study 2 Tasks: Multi-App Capabilities, Scoring Points Count, and Criteria.” It lists 8 multi-app tasks, each with a capability category, an SPC value, and scoring criteria. Tasks 1–4 are Summarization: Task 1 (SPC 6) asks for popular songs by King Gnu, listing at least six songs with details; Task 2 (SPC 8) compiles Orlando Disney travel guides from Ctrip, Zhihu, and Rednote, covering four aspects-ticket purchase, theme parks, transportation, and accommodation-with two suggestions per aspect on each platform; Task 3 (SPC 6) summarizes fan sentiment about possible future Mission: Impossible films by providing opinions from at least two platforms with three points each; Task 4 (SPC 2) reports the latest emails in Gmail and NetEase Mail by correctly retelling two contents: a subscription renewal and a login confirmation. Tasks 5–6 are Comparison: Task 5 (SPC 10) decides where to order a Big Mac by comparing Meituan and Ele.me on price, sales volume, delivery time, rating, and promotions, and providing justified conclusions and suggestions; Task 6 (SPC 8) advises on purchasing a VIVO X100 Ultra from Taobao or JD by comparing price, bundled gifts, payment discounts, and membership discounts, with correct recommendations. Task 7 is Processing (SPC 2): checks showtimes for “F1: The Movie” the day after tomorrow by providing the most recent showtimes from both platforms. Task 8 is Localization (SPC 5): identifies liked playlists across Apple Music, Spotify, and QQ Music, analyzes them, and recommends likely music styles with supporting background information.}
    \caption{Study 2 Tasks: Multi-App Capabilities, Scoring Points Count, and Criteria.}
    {
    \begin{tabular}{c p{5.5cm} c c p{7cm}}
    \toprule
    \# & Task & Capabilities & SPC & Scoring Points Criteria \\
    \hline \hline
    1 & What are some popular songs by King Gnu? & Summarization & 6 & List at least 6 songs with their information. \\ \hline
    2 & Check Orlando Disney travel guides on Ctrip, Zhihu, and Rednote. & Summarization & 8 & Cover 3 platforms, and for each of the following aspects: ticket purchase, theme park, transportation, accommodation  - provide 2 suggestions each. \\ \hline
    3 & Find out how fans feel about the possibility of more \textit{Mission: Impossible} films. & Summarization & 6 & Provide opinions from at least 2 platforms, with 3 points of opinion from each platform. \\ \hline
    4 & Check my latest email in Gmail and NetEase Mail respectively. & Summarization & 2 & Correctly retell the 2 email contents: (1) subscription renewal, (2) login confirmation. \\ \hline
    5 & Ordering a Big Mac from McDonald's, Meituan or Ele.me? & Comparison & 10 & Compare the 2 platforms on: price, sales volume, delivery time, rating, and promotional offers, and provide correct conclusions and suggestions. \\ \hline
    6 & Advise on purchasing VIVO X100 Ultra from Taobao or JD. & Comparison & 8 & Compare the 2 platforms on: price, bundled gifts, payment discounts, and membership discounts, and provide correct conclusions and suggestions. \\ \hline
    7 & Check the showtimes available the day after tomorrow of \textit{the movie F1: The Movie}. & Processing & 2 & Provide the most recent showtimes from both platforms. \\ \hline
    8 & Check my liked playlists on Apple Music, Spotify, and QQ Music, and suggest music styles I’d probably like with some background info. & Localization & 5 & Identify playlists across 3 platforms, analyze them, and give music style recommendations with supporting background information. \\
    \hline
    \bottomrule
    \end{tabular}
    }
\label{tab:study2-multi-app-task-scoring-points}
\end{table*}

\subsection{Study 3 Questions for Interviews} \label{appendix:study3-questionnaire}

\begin{enumerate}
    \item Compared with the baseline systems (manual, LLM search engines, Claude Computer Use, Mobile-Agent-v2), how do you perceive the differences in terms of task completion, take over and transparency experience?
    
    \item Do you believe the information provided by the system is accurate? In what aspects would you be concerned about potential inaccuracies? What do you usually do when you encounter such doubts?
    
    \item Do you think the information provided by the system was sufficient to support your needs?
    
    \item When using the system for information seeking, did you feel that it provided too much irrelevant or secondary information?
    
    \item Do you trust the objectivity of the system’s information retrieval and report generation? Do you think the report generation deliberately omits certain content?
    
    \item Did the system interface design help you complete tasks more efficiently? Were there any elements that you found particularly clear or confusing?
    
    \item When facing the task progress dashboard, were you able to understand what the agent was doing and how far along it was? How do you usually interpret and use this interface?
    
    \item We provided several features: risk detection with pause reminders, task progress visualization, voice input, interrupt and takeover at any time, and report generation with citation and highlighted snapshots. Which of these features did you find most useful or relatively useless? In which scenarios would you enable them?
    
    \item When using these features, did you feel you gained stronger control over the task? Which mechanisms most reassured you to delegate the task to the system?
    
    \item What types of actions or information fall within your privacy boundary, which you prefer the system not to touch?
    
    \item If these functions were developed into a formal product, in what everyday scenarios would you use it? Could you give specific examples?
\end{enumerate}


\end{document}